\documentclass[10pt,twocolumn,preprintnumbers,amsmath,amssymb,nofootinbib
,superscriptaddress]{revtex4-1}
\usepackage{graphicx,longtable,mathrsfs,color,array}
\usepackage[hidelinks]{hyperref}
\usepackage[usenames,dvipsnames]{xcolor} 
\usepackage{amssymb,amsmath,mathtools,mathrsfs,slashed} 
\usepackage{epsfig,subfigure,placeins,float} 
\usepackage{booktabs,longtable,ctable,multirow} 
\usepackage{exscale,relsize} 
\usepackage[normalem]{ulem} 
\usepackage{enumerate}
\usepackage{times, mathptmx} 
\usepackage[utf8]{inputenc}
\usepackage{color}
\usepackage{graphicx}
\usepackage{graphicx,graphics}
\usepackage{bm}
\usepackage{tabularx}
\usepackage{rotating}
\usepackage{units}
\allowdisplaybreaks[1]
\usepackage{hhline}
\usepackage{soul}

\newcommand{\nn}{\nonumber}

\newcommand{\MPl}{M_{\rm P}}

\begin{document}

\title{Theoretical priors in scalar-tensor cosmologies: Shift-symmetric Horndeski models}
\author{Dina Traykova}
 \affiliation{University of Oxford, Denys Wilkinson Building,
  Keble Road, Oxford, OX1 3RH,  UK}
\author{Emilio Bellini}
\affiliation{Département de Physique Théorique, Université de Genève, 24 quai Ernest Ansermet, 1211 Genève 4, Switzerland}
\affiliation{University of Oxford, Denys Wilkinson Building,
  Keble Road, Oxford, OX1 3RH,  UK}
\author{Pedro G.~Ferreira}
  \affiliation{University of Oxford, Denys Wilkinson Building,
  Keble Road, Oxford, OX1 3RH,  UK}
  \author{Carlos García-García}
  \affiliation{University of Oxford, Denys Wilkinson Building,
  Keble Road, Oxford, OX1 3RH,  UK}
\affiliation{Instituto de Física Fundamental, Consejo Superior de Investigaciones
Científicas, c/. Serrano 121, E–28006, Madrid, Spain}
\affiliation{ Institut de Ci\`{e}ncies del Cosmos (ICCUB), c/. Martí i Franqués 1, E–08028, Barcelona, Spain}
\author{Johannes Noller}
\affiliation{Institute of Cosmology \& Gravitation, University of Portsmouth, Portsmouth, PO1 3FX, UK}
\affiliation{DAMTP, University of Cambridge, Wilberforce Road, Cambridge, CB3 0WA, UK}
\author{Miguel Zumalac\'arregui}
\affiliation{Max Planck Institute for Gravitational Physics (Albert Einstein Institute) \\
Am Mühlenberg 1, D-14476 Potsdam-Golm, Germany}
\affiliation{Berkeley Center for Cosmological Physics, LBNL and University of California at Berkeley, \\
Berkeley, California 94720, USA}


\begin{abstract}
Attempts at constraining theories of late time accelerated expansion often assume broad priors for the parameters in their phenomenological description.
Focusing on shift-symmetric scalar-tensor theories with standard gravitational wave speed, we show how a more careful analysis of their dynamical evolution leads to much narrower priors.
In doing so, we propose a simple and accurate parametrisation of these theories, capturing the redshift dependence of the equation of state, $w(z)$, and the kinetic braiding parameter, $\alpha_{\rm B}(z)$,  with only two parameters each, and derive their statistical distribution (a.k.a. theoretical priors) that fit the cosmology of the underlying model.
We have considered two versions of the shift-symmetric model, one where the energy density of dark energy is given solely by the scalar field, and another where it also has a contribution from the cosmological constant.
By including current data, we show how theoretical priors can be used to improve constraints by up to an order of magnitude. 
Moreover, we show that shift-symmetric theories without a cosmological constant are observationally viable.
We work up to quartic order in first derivatives of the scalar in the action and our results suggest this truncation is a good approximation to more general shift-symmetric theories.
This work establishes an actionable link between phenomenological parameterisations and Lagrangian-based theories, the two main approaches to test cosmological gravity and cosmic acceleration.
\end{abstract}

\maketitle

\section{Introduction} \label{intro}
There is some hope that the evidence of accelerated expansion \cite{Riess:1998cb,Perlmutter:1998np,Abbott:2018wog,Hinshaw:2012aka,Aghanim:2018eyx,Alam:2016hwk} is an indication that new physics is at play on cosmological scales.
Thus, by characterising the evolution of the Universe in detail \cite{Abell:2009aa,Font-Ribera:2013rwa,Spergel:2013tha,Aghamousa:2016zmz,Hounsell:2017ejq}, it should be possible to measure and constrain physical parameters that capture this novel behaviour.
Typically, the new physics associated with these parameters involves new fields, a notable example of which is the scalar field, $\phi$. Indeed, shortly after the accelerated expansion was discovered, quintessence -- a scalar field whose dynamics is dominated by its potential energy -- was proposed~\cite{Ratra:1987rm,Wetterich:1987fm,Ferreira:1997hj,Caldwell:1997ii} (see also \cite{Copeland:2006wr,Tsujikawa:2013fta} for reviews).
The impact of the scalar field can be neatly encapsulated in terms of one free function, its equation of state, $w(a)$, given by
\begin{eqnarray}
w(a)\equiv\frac{P_\phi}{\rho_\phi}\,,
\end{eqnarray}
where $a$ is the scale factor and $P_\phi$ ($\rho_\phi$) are the pressure (energy density) of the scalar field.

Quintessence is part of a much larger class of theories -- scalar-tensor gravity (see \cite{Bergmann:1968ve, fujii_maeda_2003,amendola_tsujikawa_2010, Clifton:2011jh} for a review on scalar-tensor theories of gravity) -- which involves a host of possible couplings of the scalar field, both with itself and the metric.
The Horndeski family of models~\cite{Horndeski:1974wa,Deffayet:2011gz,Kobayashi:2011nu}, which leads to second order equations of motion, can be further generalised to what seems like an infinite tower of possible theories~\cite{Gleyzes:2013ooa,Zumalacarregui:2013pma}.
In principle, it should be possible to constrain such theories with observations, pinning down the fundamental parameters that enter the action.
However, given the generality of the construction, the prospects are daunting.

It turns out that it is possible to completely characterise a broad class of scalar-tensor on cosmological scales in terms of a handful of time dependent functions, $\alpha_X(a)$ (as well as $w(a)$) \cite{Bellini:2014fua, Gleyzes:2014qga} where, in the case of Horndeski gravity, $X\in\{M,K,B,T\}$ each associated to a particular physical feature of the underlying action~\cite{Bellini:2014fua}. 
A particular Horndeski model can be associated with a choice of $w$ and $\alpha_X$.
In this way, the exercise of constraining scalar-tensor gravity, reduces to finding constraints on these free functions. 
There have been a number of attempts at constraining these functions but current uncertainties are at around the $10$ to $50\%$ level~\cite{BelliniParam, Kreisch:2017uet, Mancini:2018qtb, Reischke:2018ooh, Noller:2018wyv, Noller:2018eht, SpurioMancini:2019rxy, Frusciante:2019xia, Arai:2019zul, Noller:2020afd, Baker:2020apq,Mastrogiovanni:2020gua} (see also related forecasts \cite{Gleyzes:2015rua, Alonso:2016suf}).

The typical approach for models that use phenomenological functions such as $w(a)$ and the $\alpha_X(a)$ is to assume a parametric form for their evolution and constrain its parameters.
The favoured model for $w$ is the Chevallier-Polarski-Linder (CPL) parametrisation, expansion in terms of the scale factor with coefficients $w_0$ and $w_a$ \cite{Chevallier:2000qy,Linder:2002et}.
There exist a number of well-motivated parametrisations of $\alpha_X(a)$ that assume these functions scale in some way with the fractional density parameter of dark energy (DE), $\Omega_{\rm DE}$, or the scale factor, $a$ (see e.g \cite{Bellini:2014fua,BelliniParam,Alonso:2016suf,Noller:2018wyv,Linder:2015rcz,Linder:2016wqw,Denissenya:2018mqs,Lombriser:2018olq,Gleyzes:2017kpi}).
However, what is often overlooked by making such a choice is that there are underlying physical models which may limit the ranges (and behaviours) of these functions.
One way of putting this is that the underlying physical model will impose quite strict physical priors on these functions and these should be taken into account when undertaking parameter constraints with cosmological data.
This situation is entirely analogous to what happens when constraining inflationary models.
While it is the norm to find constraint on the spectral index, $n$, and the tensor to scalar ratio, $r$, each class of inflationary models singles out very specific (often one dimensional) locii in the ($n$,$r$) plane~\cite{Bennett:2003bz,Barger:2003ym,Planck:2013jfk,Ade:2015lrj,Akrami:2018odb}.

There have been a number of studies where the evolution of the DE equation of state has been reconstructed non-parametricaly (in redshift bins) using data \cite{Sahni:2006pa, holsclaw2011nonparametric, seikel2012reconstruction, Said:2013jxa, Wang:2018fng}, as well as ones where the impact of theoretical priors on the parameters of quintessence and more general scalar-tensor theories was considered and used to introduce correlations and minimise the number of these parameters \cite{Peirone:2017lgi, Raveri:2017qvt, Espejo:2018hxa, Frusciante:2018vht,Gerardi:2019obr}.
Using a different and complementary approach, we have tackled this problem of physical priors in the case of thawing quintessence where, remarkably, we could construct an analytic prior for $w(a)$~\cite{Garcia-Garcia:2019cvr}. 
By parametrising it as 
\begin{eqnarray}
  w = w_0+w_a(1-a)\,, \label{eq:w0wa}
\end{eqnarray}
we found that if $\{w_0, w_a\}$ were chosen to fit the observables, those could be reproduced with the accuracy required by next-generation surveys up to recombination. 
Furthermore, the prior, ${\cal P}$, was factorisable, ${\cal P}[w_0,w_a]={\cal P}[w_a|w_0]{\cal P}[w_0]$ and the shape of ${\cal P}$ was such that it was not collinear with current constraints on $\{w_0, w_a\}$ and thus, if incorporated could reduce the uncertainties in $w$ by up to an order of magnitude.

Emboldened by what we have found in the case of thawing quintessence, we now wish to generalise this approach to more general scalar-tensor theories.
From the outset, it is a somewhat challenging task to construct a multidimensional probability distribution function for $w$ and $\alpha_X$.
We have therefore established a more modest goal and focused on a sub-class of theories that are shift-symmetric, i.e. theories which are invariant under a scalar field transformation of the form
\begin{eqnarray}
\phi\rightarrow \phi+C\,,
\end{eqnarray}
where $C$ is a constant.
Such theories are, in a sense we will make more precise below, well-defined and natural.
In this case, the theory is completely determined by $w(a)$, $\alpha_{\rm B}(a)$ and $\alpha_{\rm K}(a)$;
however, it is well known that $\alpha_{\rm K}(a)$ is unconstrained by observations~\cite{BelliniParam}, so we are seeking a prior distribution function for $w(a)$ and $\alpha_{\rm B}(a)$.
As we will see, exploring this restricted set of scalar tensor models already sheds light on the hurdles we need to tackle in the general case. 
Note that, motivated by recent observations \cite{PhysRevLett.119.161101,2041-8205-848-2-L14,2041-8205-848-2-L15} and associated theoretical bounds \cite{Baker:2017hug, Ezquiaga:2017ekz,Creminelli:2017sry,Sakstein:2017xjx}, in the above we have implicitly required that the speed of gravitational waves is luminal.

{\it Outline:} In Section~\ref{shift} we outline the theoretical aspects of and motivation for the shift-symmetric Horndeski model that we focus on here.
In Section~\ref{priors} we justify the choice of physical priors we impose on the theory.
In Section~\ref{approx} we describe the approximation scheme we use here and explain how we evaluated the required accuracy.
Further, in Section~\ref{results} we present the constructed prior functions on $w$ and $\alpha_{\rm B}$.
In Section~\ref{data} we combine these priors with a set of cosmological data.
Finally in Section~\ref{disc} we discuss our findings.

\section{Shift-Symmetric Scalar-Tensor Gravity}
\label{shift}
Consider as a starting point, the Horndeski action \cite{Horndeski:1974wa,Deffayet:2011gz,Kobayashi:2011nu}: 
\begin{equation}
S[g_{\mu\nu},\phi]=\int\mathrm{d}^{4}x\,\sqrt{-g}\left[\sum_{i=2}^{5}\frac{1}{8\pi G_{\text{N}}}{\cal L}_{i}[g_{\mu\nu},\phi]\,+\mathcal{L}_{\text{m}}[g_{\mu\nu},\psi_{M}]\right]\,,\label{eq:action}
\end{equation}
where $\mathcal{L}_{\text{m}}$ captures the matter Lagrangian, with all matter fields $\psi_M$ minimally coupled to $g_{\mu\nu}$ (in other words, we are in the Jordan frame), and where
\begin{eqnarray}
{\cal L}_{2} & = & G_{2}(\phi,\, X)\,,\label{eq:L2}\\
{\cal L}_{3} & = & -G_{3}(\phi,\, X)\Box\phi\,,\label{eq:L3}\\
{\cal L}_{4} & = & G_{4}(\phi,\, X)R+G_{4X}(\phi,\, X)\left[\left(\Box\phi\right)^{2}-\phi_{;\mu\nu}\phi^{;\mu\nu}\right]\,,\label{eq:L4}\\
{\cal L}_{5} & = & G_{5}(\phi,\, X)G_{\mu\nu}\phi^{;\mu\nu} -\frac{1}{6}G_{5X}(\phi,\, X)\Big[\left(\Box\phi\right)^{3} \nn \\
&&+2{\phi_{;\mu}}^{\nu}{\phi_{;\nu}}^{\alpha}{\phi_{;\alpha}}^{\mu}-3\phi_{;\mu\nu}\phi^{;\mu\nu}\Box\phi\Big]\,. \label{eq:L5}
\end{eqnarray}
Here $X\equiv - \frac{1}{2}\nabla^\mu \phi\nabla_\mu\phi$, covariant derivatives on $\phi$ are denoted by indices, so e.g. $\phi_{;\mu}{}^{\nu} \equiv \nabla_\mu\nabla^\nu\phi$, and similarly we use a shorthand for partial derivatives wrt. $X$, e.g. $G_{4X}=\partial G_4/\partial X$.
The Horndeski action describes the most general Lorentz invariant, local action in four dimensions, featuring a scalar field on top of the metric and having at most second-order equations of motion on any background.
Even if the final aim, beyond the scope of this paper, is to investigate the impact of physical priors for this action in full generality, in this paper we focus on a simpler scenario: {\it shift-symmetric} Horndeski theories.
This subset of theories is also known as `weakly broken Galileons' \cite{Pirtskhalava:2015nla}, since the shift symmetry ensures that radiative corrections are parametrically suppressed around (quasi) de Sitter backgrounds, reminiscent of non-renormalisation theorems for Galileons \cite{Luty:2003vm,Nicolis:2008in}.\footnote{Although see \cite{Noller:2018eht,Heisenberg:2020cyi} for examples of shift-symmetry breaking theories that maintain this property.}
By focusing on this subset of solutions we are therefore already implicitly ensuring that a theoretical prior requiring the radiative stability of the theory is satisfied.\footnote{By this we mean radiative stability of the Horndeski scalar interactions considered here. We have nothing new to say about the old cosmological constant problem.}

As we are ultimately interested in investigating concrete cosmological observables for shift-symmetric Horndeski theories (and the effect theoretical priors have on them), we need to choose a concrete parametrisation of the (in principle infinite) freedom inherent in the $G_i$ functions.
As a concrete illustration we therefore focus on the following subset of theories
\begin{align} 
G_{2}&= c_{01} X + \frac{c_{02}}{\Lambda_2^4} X^2, &G_{3} &= -\frac{1}{\Lambda_3^3}(d_{01} X + \frac{d_{02}}{\Lambda_2^4} X^2)\,,\nn\\
G_{4}&= \tfrac{1}{2}M_{P}^{2}, &G_{5} &= 0\,.
\label{eq:Gexample}
\end{align}
belonging to the Kinetic Gravity Braiding (KGB) \cite{Deffayet:2010qz} class.
Here the reduced Planck mass is $M^2_P=1/8\pi G$ and conventionally $\Lambda_2^4 = \MPl^2 H_0^2,\, \Lambda_3^3 = \MPl H_0^2$, ensuring all the above interactions can give ${\cal O}(1)$ contributions to the cosmological background evolution today.
The choice for $G_4$ and $G_5$ is dictated by constraints on the speed of gravitational waves \cite{PhysRevLett.119.161101,2041-8205-848-2-L14,2041-8205-848-2-L15,2041-8205-848-2-L13,2041-8205-848-2-L12} -- see \cite{Creminelli:2017sry,Sakstein:2017xjx,Ezquiaga:2017ekz,Baker:2017hug} and references therein for why this implies the above restrictions on the $G_i$, at least as long as the cosmological Horndeski theory is valid up to energy scales of $\Lambda_3$ \cite{deRham:2018red}.
For $G_{2,3}$ we keep the first two orders in $X$, where the $c_{01}$ and $d_{01}$ terms capture the Galileon symmetric contributions, while the $c_{02}$ and $d_{02}$ capture the lowest order (in $X$) shift-symmetric corrections to this.\footnote{If higher order terms in $X/\Lambda_2^4$ are suppressed (while terms such as $(\Box\phi)^n/\Lambda_3^{3n}$ are not), then this will fully capture the leading order terms as well as next-to-leading-order corrections for a generic $G_{2,3}$. If higher-order terms are not suppressed and e.g. all powers of $X/\Lambda_2^4$ equally contribute to $G_{2,3}$, this is not the case.
A truncation like Eq.~\eqref{eq:Gexample} is therefore not generically valid, but instead it should be viewed as a specific illustrative example of a shift-symmetric Horndeski theory.}
This will afford us with a fairly minimal, yet suitably rich testbed in which to investigate the effect of theoretical priors on shift-symmetric Horndeski theories.
Note that, for simplicity, we have excluded the (equally shift-symmetric) tadpole term $c_{10} \phi$ in our test case, Eq.\eqref{eq:Gexample}. 

The shift-symmetric model has been explored previously in Refs.~\cite{Peirone:2019aua} and \cite{Frusciante:2019puu}, where the authors put cosmological constraints on the parameters of the model, defined in Eq.~\eqref{eq:Gexample}, and on the parameters of the shift-symmetric generalisation of the Cubic Covariant Galileon model, respectively.

We will be considering a homogeneous and isotropic cosmological (FRW) background solution, $ds^2=-dt^2+a^2(t)(d{\bf x})^2$, populated by matter, radiation and the dark energy scalar $\phi$.
The Friedmann equations then are 
\begin{align}
H^2 &= \frac{1}{3\MPl^2}\rho_{\rm tot}\,, 
&\dot H &= -\frac{1}{2\MPl^2}\left(\rho_{\rm tot}+p_{\rm tot}\right)\,, 
\label{eq:frw}
\end{align}
where $H \equiv {\dot a}/a$ as usual, $\rho_{\rm tot} = \rho_m + \rho_r + \rho_{\phi}$ and $p_{\rm tot} = p_r + p_{\phi}$ (subscripts refer to matter, radiation and dark energy, respectively). For Eq.~\eqref{eq:Gexample}, $\rho_{\rm DE}$ and $p_{\rm DE}$ then satisfy
\begin{align}
\rho_{\phi} &= \frac{1}{2}\left(c_{01}+\frac{3}{2}\frac{c_{02}}{\Lambda_2^4}\dot\phi^2\right)\dot\phi^2 - \frac{3}{\Lambda_3^3} \left(d_{01} + \frac{d_{02}}{\Lambda_2^4}\dot\phi^2\right) H \dot\phi^3,\nn \\
p_{\phi} &= \frac{1}{2}\left(c_{01}+\frac{1}{2}\frac{c_{02}}{\Lambda_2^4}\dot\phi^2\right)\dot\phi^2 + \frac{1}{\Lambda_3^3}\left(d_{01} + \frac{d_{02}}{\Lambda_2^4}\dot\phi^2\right) \dot\phi^2\ddot\phi\,.
\label{eq:rho_p}
\end{align}
Note that in the case where we include a cosmological constant $\Lambda$, described in more detail below, the dark energy density, $\rho_{\rm DE}$, and pressure, $P_{\rm DE}$, will have a contribution from $\Lambda$ in addition to $\phi$. However, in both cases we take $w(a)=P_\phi/\rho_\phi$.

The background scalar equation of motion can be written in terms of a conserved current \cite{Bellini:2014fua} as
\begin{align}
\dot J  + 3HJ = 0\,,
\label{eq:Jeq}
\end{align}
where
\begin{align}
J &= \left(c_{01}+\frac{c_{02}}{\Lambda_2^4}\dot\phi^2\right)\dot\phi - \frac{3}{\Lambda_3^3} \left(d_{01} + \frac{d_{02}}{\Lambda_2^4}\dot\phi^2\right) H \dot\phi^2\,.
\label{eq:Jdef}
\end{align}

There are a few key points to note about the background evolution. 
First of all, we have that Eq.~\eqref{eq:Jeq} implies that there is a tracker solution as $J\propto a^{-3} \rightarrow 0$ as $a$ grows.
This greatly simplifies the dynamics and, as we will reiterate further down, the priors we need to assume on the various ingredients of this model.
Second, we will consider two versions of this theory.
In the first version the scalar field is entirely responsible for the late time acceleration and thus there is no explicit cosmological constant, $\Lambda$ (or a constant term $V_0$ in the scalar field potential); we will dub this the \textit{$\Lambda=0$ self-accelerating} version.
\footnote{We mean self-acceleration in the sense that the scalar field provides accelerating expansion, i.e. $w_{\phi}<-1/3$. Note that some authors use the term self-acceleration to mean that only the Jordan-frame scale factor is accelerating (while its Einstein-frame counterpart is not) \cite{Nicolis:2008in}. This can not be the case in the theory at hand, as both frames are equivalent.}
The $\Lambda=0$ version is, in some sense, the more interesting as it can be invoked as an alternative to cosmological constant driven acceleration.
But we also have experience from other theories that self-accelerating solutions are more tightly constrained and potentially easier to rule out (for example in the case of GDP gravity \cite{Dvali_2000, Deffayet_2001, Fang_2008, Schmidt_2009}).
This means that the dark energy density is solely given in terms of the energy density associated to the scalar field: $\Omega_{\rm DE}=\Omega_\phi$.

\begin{figure}[t!]
  \centering
   \includegraphics[trim={0.3cm 0.3cm 0 0},clip,width=0.48\textwidth] {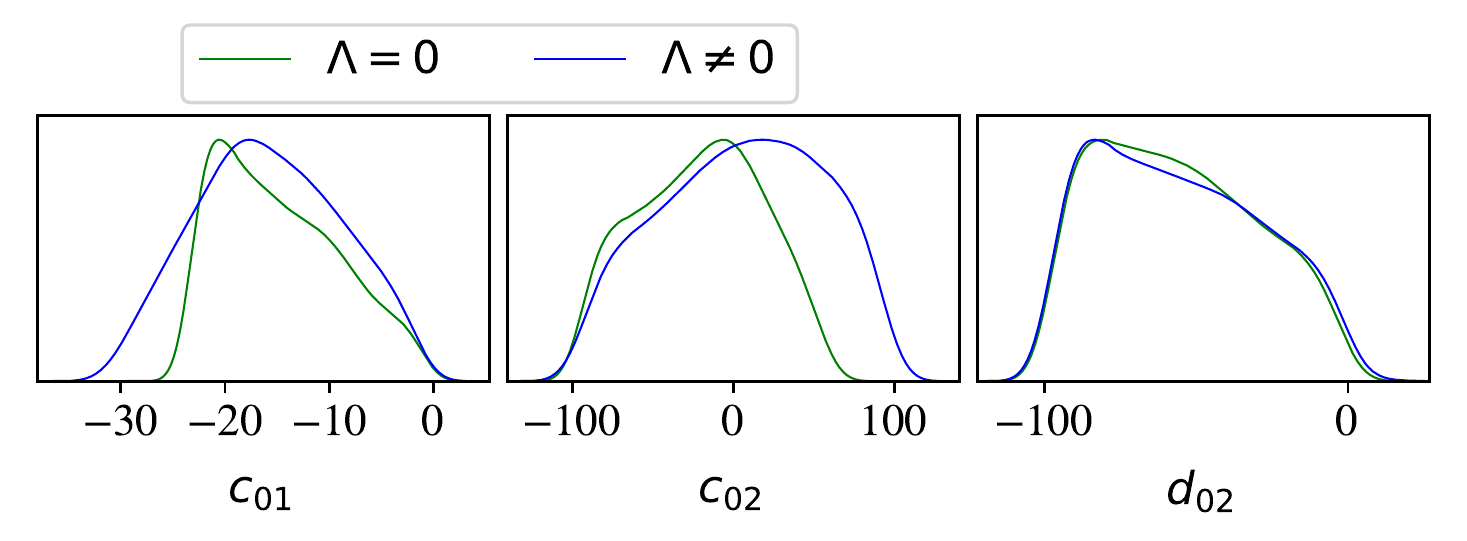}
   \caption{Distributions of the parameters of the action for the $\Lambda=0$ (green) and $\Lambda\neq0$ (blue) variants of the shift-symmetric model, where we have fixed $d_{01}=-1$. $\Lambda=0$ is the first version of shift-symmetric theories we consider, where there is no explicit cosmological constant, $\Lambda$, and the density of DE is given solely by the scalar field $\phi$; in the case of $\Lambda\neq0$, $\Omega_{DE}$ has contributions both from $\phi$ and $\Lambda$.}
  \label{fig:ci_normal-vs-self-accel}
\end{figure}

A key aspect of self-accelerating solutions is that they require ``negative kinetic energy'' $G_2<0$, at least in the class of theories under consideration \cite{Deffayet:2010qz}. For shift-symmetric Horndeski theories up to cubic term (Kinetic Gravity Braiding), the energy density can be written as \cite{Deffayet:2010qz}
\begin{equation}\label{eq:kgb_energy_density}
  \rho_{\phi} = \dot\phi J - G_2 \to -G_2\,,
\end{equation}
where the latest limit corresponds to the tracker solution.
Because $G_2$ is even in $\dot\phi$, $\rho_{\phi}>0$ requires that at least one of $c_{01},c_{02}$ to be negative (the tracker condition $J=0$ on Eq.~\eqref{eq:Jdef} might impose further constraints on the relative signs).
We will find that generically $c_{01}<0$, i.e. the ``wrong'' sign of the standard kinetic term, Fig.~\ref{fig:ci_normal-vs-self-accel}.
\begin{figure}[t!]
  \centering
   \includegraphics[width=0.48\textwidth]
   {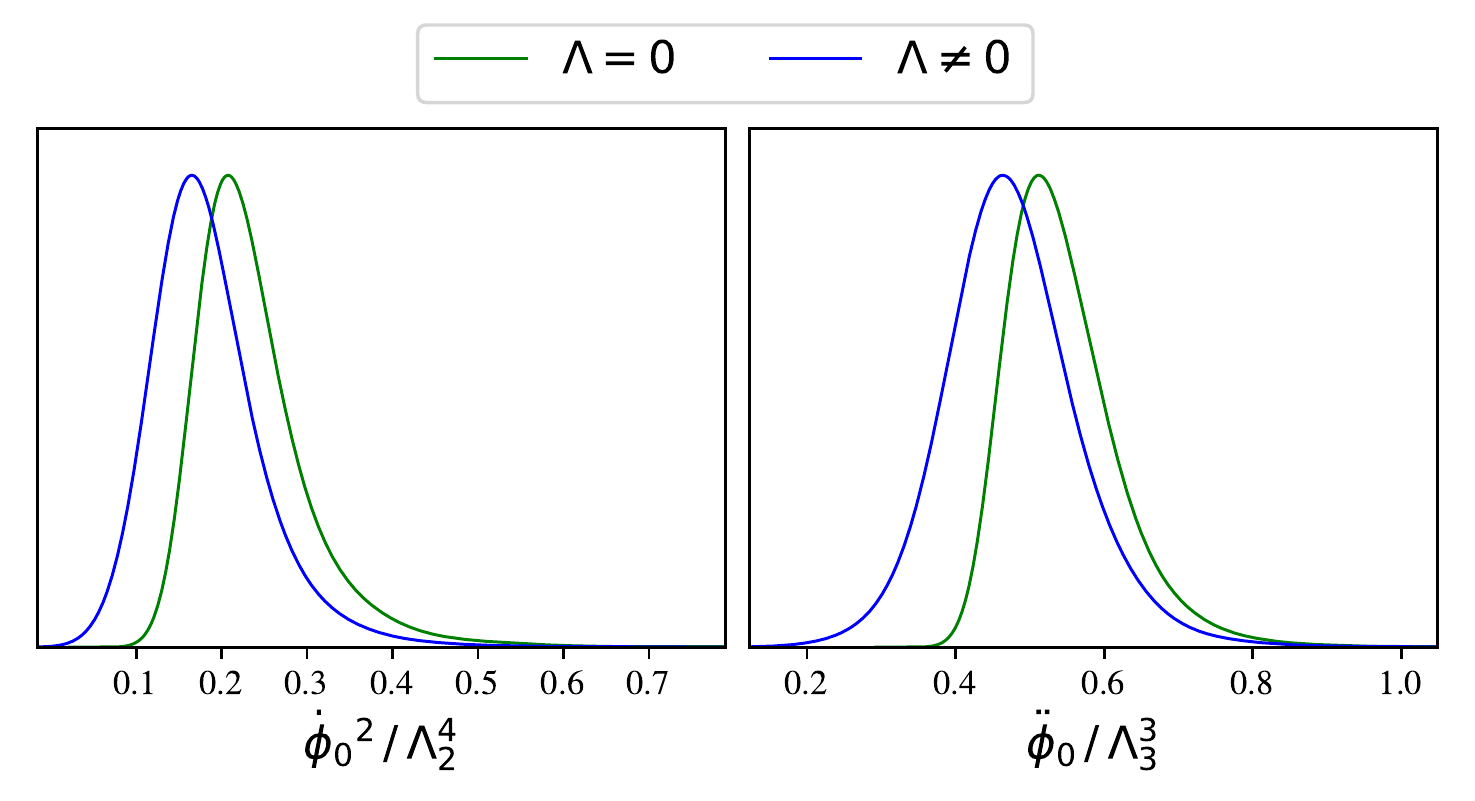}
   \caption{Distributions of the parameters of  ${\dot\phi_0}^2/\Lambda_2^2$ and $\ddot\phi_0/\Lambda_3^3$, where $\phi_0$ is the amplitude of the scalar field today. The fact that they are lower than 1 means that higher order terms in our expansion of the Lagrangian, Eq.~\eqref{eq:Gexample}, should be suppressed unless large values of the coefficients $\{c_{ij}, d_{ij}\}$ were chosen. Therefore, this could be seen as a posteriori justification of our ansatz, Eq.~\eqref{eq:Gexample}.}
  \label{fig:dotphi}
\end{figure}
This means that Minkowski space with $\dot\phi=0$ is not a stable solution of these models nor can we apply the usual battery of consistency conditions that have been developed in the standard vacuum (see discussion in the next section).
\footnote{The theories under consideration have some other generic properties: for instance the equation of state is phantom $w_\phi<-1$ in the tracker, approaching de Sitter $w_\phi\to -1$ from below as $\rho_m\to 0$ \cite{Deffayet:2010qz}.}

Another interesting feature of the self-accelerating solutions is illustrated in Fig.~\ref{fig:dotphi}. 
There we can see that both $\dot\phi_0^2/\Lambda_2^4$ and $\ddot\phi_0/\Lambda_3^3$ are smaller than unity. 
This is encouraging in that it provides a posteriori justification for our ansatz, Eq.~\eqref{eq:Gexample}: the higher order terms omitted in Eq.~\eqref{eq:Gexample} scale with higher powers of $\dot\phi_0^2/\Lambda_2^4$ and $\ddot\phi_0/\Lambda_3^3$. 
So if these higher powers are indeed suppressed, then omitting higher order terms in the first place is consistent. 
This is also related to the above discussion of the sign of $c_{01}$. 
If higher order terms with coefficients $c_{0i}$ and $i > 1$ are increasingly suppressed, then obtaining a positive scalar energy density, Eq.~\eqref{eq:kgb_energy_density}, with positive $c_{01}$ becomes very challenging.
Note, however, that the suppression illustrated in Fig.~\ref{fig:dotphi} is rather mild and can easily be compensated for by coefficients $c_{ij}$ and $d_{ij}$ that are somewhat larger than unity. 
Fig.~\ref{fig:ci_normal-vs-self-accel} shows that this is in fact the case for the lower order interactions in our ansatz, Eq.~\eqref{eq:Gexample}, so we emphasise that our findings here are certainly not conclusive evidence that the higher order interactions omitted cannot yield ${\cal O}(1)$ contributions to the scalar energy density or the background and perturbative evolutions in general. 

The second variant that we will consider does include $\Lambda$; we will dub it the $\Lambda\neq 0$ version. In this case the signs of  $c_{01},\,c_{02}$ are less restricted by requiring the scalar field to dominate the expansion, Eq.~\eqref{eq:kgb_energy_density}. 
If we were to restrict ourselves to $c_{01}>0$ (which we do not here) we would be looking at what is conventionally dubbed the normal branch.
In the cubic Galileon limit ($c_{02},\,d_{02}=0$) $\Omega_{\phi}>0$ requires $c_{01}<0$, in agreement with Eq.~\eqref{eq:kgb_energy_density}. 
Normal-branch Galileons ($c_{01}>0$) are driven towards a trivial tracker with $\dot\phi\to 0$, $\rho_{\phi}\to 0$ unless shift-symmetry is broken \cite{Zumalacarregui:2020cjh}.
We will not fix a sign of $c_{02}$ to be able to capture more general behaviour in the $\Lambda\neq0$ case.
Note that the cosmological constant is allowed and does not break shift symmetry. Here the dark energy density is the sum of the energy density associated to the scalar field and the cosmological constant: $\Omega_{\rm DE}=\Omega_\phi+\Omega_\Lambda$. 

As we will focus on large scale observables, we are particularly interested in linearised perturbations around the cosmological background solution described above.
The freedom in the dynamics of such perturbations for a general Horndeski theory, as specified in Eqs.~\eqref{eq:action}--\eqref{eq:L5}, is controlled by just four functions $\alpha_X$ of time with $X \in \{K, B, M, T\}$. For the general form of these $\alpha_X$ see \cite{Bellini:2014fua}.
In the shift-symmetric subset of theories we are considering here, with $G_{4X} = 0 = G_{5X}$, we find that the effective Planck mass seen by linear perturbations is simply $\MPl$ (and hence has no time-dependence), while the speed of gravitational waves $c_{\rm GW} = 1$ by construction. 
We are therefore left with only two non-trivial $\alpha_X$ controlling linear perturbations, namely
\begin{align}
H^2M^2_P{\alpha}_{\rm K} &=  2X \left[G_{2X}+2XG_{2XX} +6\dot{\phi}H\left(G_{3X}+XG_{3XX}\right)\right], \nonumber \\
H^2M^2_P{\alpha}_{\rm B} &= 2X\dot{\phi}HG_{3X},
\label{eq:alphas}
\end{align}
where all functions are evaluated at the background level.
Upon substituting Eq.~\eqref{eq:Gexample} into Eq.~\eqref{eq:alphas}, it is then straightforward to express these two $\alpha_X$ in terms of the $c_{ij}, d_{ij}$ in Eq.~\eqref{eq:Gexample} and the background degrees of freedom, $a$ and $\phi$.

\section{Establishing Physical Priors}
\label{priors}

It has been well established that cosmological observables are insensitive to $\alpha_{\rm K}$ \cite{BelliniParam},  a direct manifestation of the fact that $\alpha_{\rm K}$ drops out in the quasi-static limit (which applies to the vast majority of observable scales at late times) at leading order \cite{Alonso:2016suf}.
The challenge, then, is to construct physical priors for $w$ and $\alpha_{\rm B}$.
There are a number of steps in working towards this goal, the first one of which is to map out the space of possible histories for the scalar field $\phi$ and the metric $g_{\alpha\beta}$.
In fact, as we saw in the previous section, $w$ and $\alpha_{\rm B}$ are completely determined in terms of $\phi(t)$ and $a(t)$ so we will only have to focus on the evolution of the background in these theories.

We then have a number of parameters which need to be chosen.
The standard cosmological parameters will be included in the analysis, whether we work with the scalar field action directly or we work with the parametrised form, in terms of $w$ and $\alpha_{\rm B}$; therefore, we will not be specially concerned with the choice of their priors; indeed we will consider a standard range such as  $\Omega_{\rm cdm}\in\left[0.15, 0.35\right]$ and $H_0\in[60,80]\,\unit[]{km\,s^{-1}\,Mpc^{-1}}$, which ensures our findings will be compatible with current constraints of these parameters, while not too broad to explore values that are already ruled (e.g. $H_0 = 0$).
We then have the parameters in the action which we have distilled down to $\{c_{01},c_{02},d_{01},d_{02}\}$ and $\{\Lambda_2.\Lambda_3\}$. 
Two dimensionless $\left\{c_{ij}, d_{ij}\right\}$ can be absorbed into the $\Lambda_i$. 
In our concrete implementation, however, we find it more practical to follow a different (yet physically equivalent) prescription and fix $\Lambda_2^4 \equiv \MPl^2 H_0^2$ and $\Lambda_3^3 \equiv \MPl  H_0^2$, varying only the coefficients $\{c_{01},\,c_{02},\,d_{02}\}$.
In order to decouple the effects of $H_0$ on the coefficients and avoid possible inconsistencies due to  the way we choose to sample our parameters, we set this normalisation $H_0$ to a fiducial value.
We use the fact that we can set $d_{01}=-1$ due to the normalization of the field \cite{Barreira:2013jma}.
\footnote{Fixing the sign of $d_{0i}$ bears no loss of generality: because $L_3$ contains only odd powers of $\phi$, changing the sign of $d_{0i}$ is equivalent to flipping the sign in the initial $\dot\phi$. In contrast, normalizing the field to fix a coefficient in $L_2$ restricts the theory, cf. Eq, 11 in Ref. \cite{Zumalacarregui:2020cjh}.}
The physics of the model does not change depending on which of the $c_{0i}$ (up to its sign) and $d_{0i}$ parameters one chooses to fix, and hence our priors would be unaffected by this choice.

And finally, we must also consider the initial conditions of the scalar field, $\phi_i$ and ${\dot \phi}_i$.
As we have seen in the previous section, shift-symmetric theories come endowed with tracking behaviour.
This means that irrespective of the initial condition, the field will (quite rapidly) evolve towards a universal solution which is uniquely determined in terms of the coupling constants of the theory.
And, because the theory is shift-symmetric, the result is completely independent of $\phi_i$.
This means that the prior will also be completely independent of $\phi_i$ and ${\dot \phi}_i$.

\begin{figure}
  \centering
   \includegraphics[trim={0.3cm 0 0 0},clip,width=0.48\textwidth] {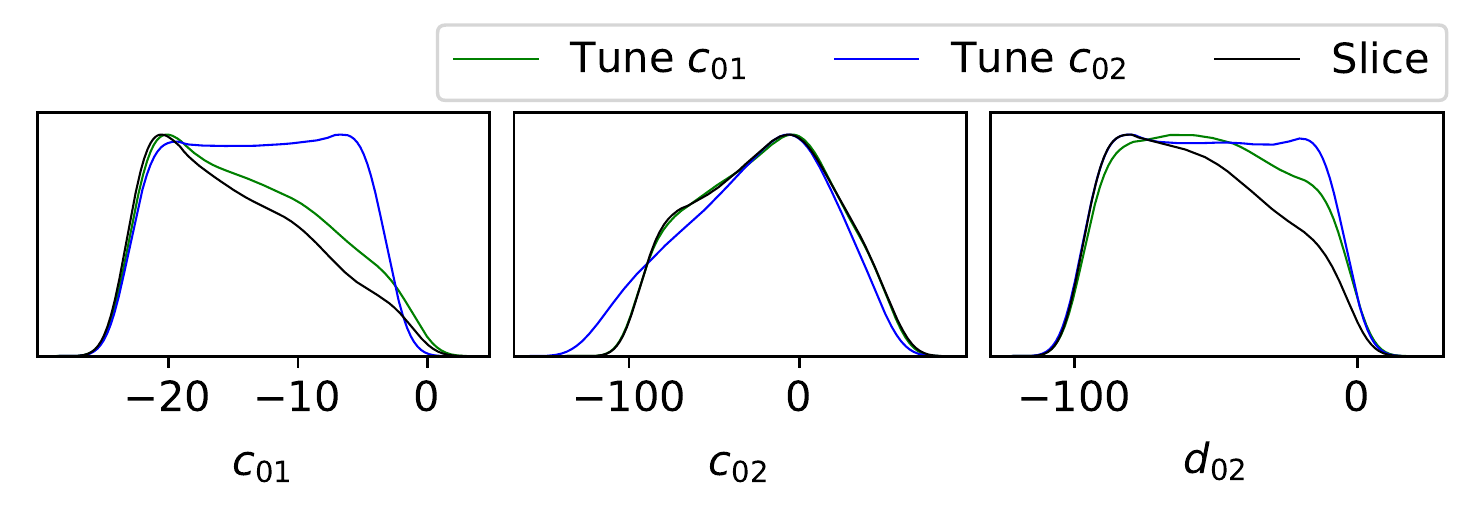}
   \caption{Distributions of the parameters given different way of sampling. In the case of \textit{tuning} one parameter, one samples the other two and chooses the value of the former that result in the desired $H_0$. In contrast, in the case of \textit{slicing}, one varies the all three parameters simultaneously and keeps only the sets that give $\sum\Omega_i = 1$. As can be seen above, the \textit{tuning} method is subject to projection effects.}
  \label{fig:tune-vs-no_tune}
\end{figure}

With regards to $\{c_{01},c_{02},d_{02}\}$, it makes sense to consider uniform, uncorrelated priors over a fixed range; as with all uniform priors, one needs to define hard limits to their ranges. 
One may expect that naturalness criteria suggest one should only vary these dimensionless constants within a range of ${\cal O}(1)$ (around 0). 
However, note that the shift-symmetric nature of the theories at hand means that radiative corrections to the $c_{0i}$ and $d_{0i}$ are parametrically suppressed \cite{Pirtskhalava:2015nla}, more specifically these corrections scale as $\left\{\delta c_{ij},\delta d_{ij}\right\} \sim \left(\Lambda_3/\Lambda_2\right)^4 \sim 10^{-40}$ \cite{Pirtskhalava:2015nla,Creminelli:2017sry}. 
So considerably wider prior ranges can be explored without running into naturalness issues. 
We have explored different choices for the ranges of these parameters and have found that, once we allow them to vary within a range of ${\cal O}(2)$, the final results are unchanged.
We also check that the constraints with data are consistent with these bounds, and indeed we find distributions that are well within the range of ${\cal O}(2)$. 
This confirms that this is a wide enough range so that our results are not biased by the bounds we have chosen, while at the same time we exclude regions of space that are ruled out by data.
We use such an extended range in all our subsequent results.

There is a further complication, however, which is that we are interested in cosmologies which are reasonably close to the one we observe, i.e. one in which $\Omega_{\rm r} + \Omega_{\rm m} + \Omega_{\rm DE} = 1$ (note that our definition of $\Omega_{\rm DE}$ differs between the cases with and without $\Lambda$); one can loosen this statement and say that we do not want $\Omega_{\rm DE}\simeq 0$ or $\Omega_{\rm DE}\simeq 1$.
This immediately imposes additional restrictions on $\{c_{01},c_{02},d_{02}\}$.
In other words, one can see such a restriction due to $\Omega_{\rm DE}$ as a deformed slab cutting through $\{c_{01},c_{02},d_{02}\}$, picking out a lower dimensional space.
Projecting such a cut onto each of the $\{c_{01},c_{02},d_{02}\}$ will naturally lead to non-uniform 1-D priors.

One might think that an alternative approach is to solve for one of the $\{c_{01},c_{02},d_{02}\}$ for a fixed range of $\Omega_{\rm DE}$ and indeed it is possible to do so using a well established shooting method.
Unfortunately the resulting combined priors depend heavily on which of the constants one chooses to solve for.
One can understand this if one takes two examples.
In one case, one assumes a uniform prior for $\{c_{02},d_{02}\}$ and solves for $c_{01}$.
The resulting distribution $c_{01}$ will not, generally be uniform.
Alternatively one might consider a uniform prior for $\{c_{01},d_{02}\}$ and solves for $c_{02}$.
Now the prior on $c_{01}$ {\it will} be uniform while the prior on $c_{02}$ will not be uniform.
We illustrate this in Fig.~\ref{fig:tune-vs-no_tune}.
This is not surprising as this approach effectively introduces a non-linear correction to the measure which is highly dependent on the constant one is solving for.
Thus we have opted to use original approach -- to sample all the parameters and then project down the constraint slice (or slab)~\footnote{{Note that, by default, \texttt{hi\_class} adjusts one of the parameters to fulfil the Friedman equation. In order to prevent this, you must set \texttt{Omega\_smg\_debug} and unset \texttt{Omega\_smg}.}}. 

A comment is in order about imposing priors related to the validity of the underlying theory itself. 
Firstly, these come in the form of stability priors. 
We have already alluded to radiative stability above and we will complement this by requiring the absence of ghost and gradient instabilities for our cosmological solution, using the implementation of \cite{Zumalacarregui:2016pph}. 
Note that these instabilities directly manifest themselves in the effective (low-energy and classical) theory we are considering, i.e. Eqs.~\eqref{eq:action}--\eqref{eq:Gexample}.\footnote{A direct consequence of this is that instability-infested regions of parameter space generically give very poor fits to the data. In other words, had we not imposed these priors, the data would still generically have excluded these regions of parameter space.} 
Secondly, there are priors not directly linked to any easily recognisable sickness in the low-energy theory, but instead to ensuring that this low-energy theory can be embedded in a sensible UV completion. 
These bounds turn out to be powerful, even if the UV completion is not known. 
In this context we will focus on so-called positivity bounds, requiring that the underlying fundamental theory (and hence the UV completion as well) is consistent with a ``standard'' Wilsonian field theory description -- one in which Lorentz invariance, unitarity (well-defined probabilities), analyticity (causality) and polynomial boundedness (locality) are respected. 
These basic principles turn out to be sufficient in order to derive a variety of additional constraints on the low energy parameters of the theory, in our case encoded in the $c_{ij}$ and $d_{ij}$ --  
see \cite{Adams:2006sv,Jenkins:2006ia,Nicolis:2009qm,Bellazzini:2014waa,Baumann:2015nta,deRham:2017avq,deRham:2017imi,deRham:2017xox,Bellazzini:2019xts, deRham:2019ctd,Ye:2019oxx,pos,Kennedy:2020ehn,Alberte:2020jsk,Tokuda:2020mlf,Bellazzini:2020cot,Tolley:2020gtv,Grall:2021xxm,deRham:2021fpu} for constraints directly applicable to our present scalar-tensor context. 
The simplest such bounds can be derived via considering tree-level $2 \to 2$ scattering on a flat (Minkowski) background. 
For general Horndeski theories the resulting bounds are presented in \cite{pos}.
Specialised to Eq.~\eqref{eq:Gexample}, these reduce to  
\begin{align}
\bar{G}_{2,XX}  &\geq 0  \quad\quad \Rightarrow \quad\quad c_{02} \geq 0\,, \nn \\
\bar{G}_{3,X}^2 &\geq 0  \quad\quad \Rightarrow \quad\quad d_{01}^2 \geq 0 \,  ,  \label{eq:posbounds}   
\end{align}
where a bar denotes that the function is evaluated on a flat background ($\langle \phi \rangle = 0$) and constraints are subsequently ported to cosmological backgrounds. 
While the second bound is trivially satisfied, the first imposes a non-trivial constraint. 
However, and crucial to the results of this paper, these bounds will turn out to {\it not} be applicable here.
This is because we will find that, for our ansatz, Eq.~\eqref{eq:Gexample}, cosmological constraints push $c_{01}$ to be overwhelmingly negative. 
While this condition is consistent with obtaining healthy solutions on cosmological backgrounds, around a flat (Minkowski) space-time it renders $\phi$ into a ghost. 
But the existence of a well-defined and ghost-free Minkowski solution is an essential ingredient for the derivation of the above positivity bounds. 
So, at least for our specific ansatz, Eq.~\eqref{eq:Gexample}, we will not be able to identify regions of parameter space here, where observational constraints are satisfied {\it and} where we can consistently apply the above positivity bounds -- for a more detailed discussion see \cite{posv2}.

\section{Approximating the time dependence of \texorpdfstring{$w$ and $\alpha_{\rm B}$}{Lg}}
\label{approx}

\begin{figure}
        \includegraphics[width=0.5\textwidth]{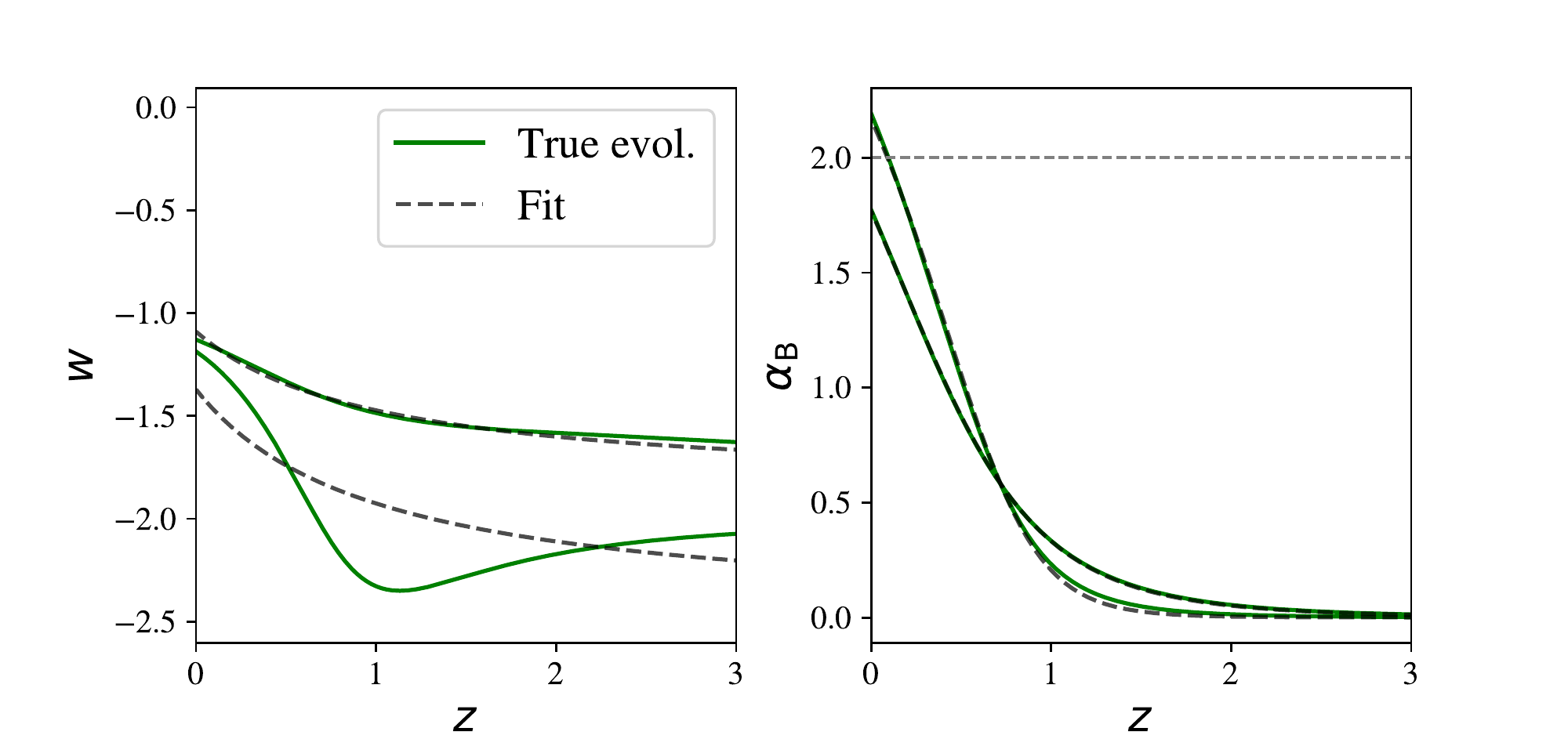}
    \caption{The time evolution of $w$ (left) and $\alpha_{\rm B}$ (right) for two representative models and their approximate fits using the parametrisations we choose. Note that we have highlighted the $\alpha_{\rm B} = 2$ line here and in the following plots, as evolutions that cross this line (as shown here $\alpha_{\rm B}$ typically increases monotonically in time) display some singular behaviour -- see Appendix~\ref{app-disco} for a more detailed discussion.}
    \label{fig:aB_w-evol}
\end{figure} 

We now proceed to determine the best way to parametrise the time dependence of $w$ and $\alpha_{\rm B}$. 
We recall that, in the case of thawing quintessence, we found that $w=w_0+w_a(1-a)$ was an excellent approximation to the equation of state; this was not the case for tracking quintessence. 
On the left panel of Fig.~\ref{fig:aB_w-evol} we plot the two typical shapes of the evolution of $w$ for the shift-symmetric model that we consider; although, on the face of it, the true curve and the fit do not seem to agree particularly well, we find that $w=w_0+w_a(1-a)$ approximates the evolution of equation of state well enough in a sense that will be clear soon.

A natural first choice for the time-dependence of $\alpha_{\rm B}$ was the commonly assumed scaling with fractional density of DE, $\Omega_{\rm DE}$, however, we found this parametrisation to only provide a good fit to a small fraction of the models we calculated. We found similar results for a proportionality with the scale factor, $a$. A Taylor expansion in terms of either $a$ or $\Omega_{\rm DE}$ worked better than a single constant factor, however, in order to reach our desired error for a the majority of models, it was necessary to include at least seven coefficients in the expansion. We had similar success with other parametrisations, such as inverse power law, binomial expansion, exponential power law and others.
Finally, we note that in a simplified version of our model, i.e.~the cubic galileon,\footnote{The cubic galileon model is equivalent to a special case of the shift-symmetric we consider here with $c_{02} = d_{02} = 0$.} its exact time dependence is $\alpha_{\rm B} \propto H^{-4}$.
The tracker solution, $J=0$ in Eq.~\eqref{eq:Jdef}, provides a solution for the scalar field evolution $\dot{\phi}\propto H^{-1}$, which can be substituted into Eq.~\eqref{eq:alphas} to get the expected result.
It is also possible to prove that this time-dependence approximately holds for a more general case too, i.e.~$c_{02} \neq 0$ and $d_{02} = 0$.
Therefore we expect that a function of $(H_0 / H)^4$ should fit the evolution of $\alpha_{\rm B}$ in the shift-symmetric case.
We find the following function to fit the true models extremely well,
\begin{equation}
	\alpha_{\rm B} = \hat{\alpha}_{\rm B}\left(\frac{H_0}{H}\right)^{4/m} \,,
    \label{eq:aB-param}
\end{equation}
where $\hat{\alpha}_{\rm B}$ and $m$ are constant parameters.
On the right of Fig.~\ref{fig:aB_w-evol} we plot two typical representative $\alpha_{\rm B}$ and the lines that fit to those given the function that we chose.
As will be discussed in more detail in the following section~\ref{results} this parametrisation fits incredibly well (to less than $1\%$ error) more than $98\%$ of the large set of randomly generated models.
\begin{figure}[t!]
  \centering
   \includegraphics[width=0.5\textwidth]{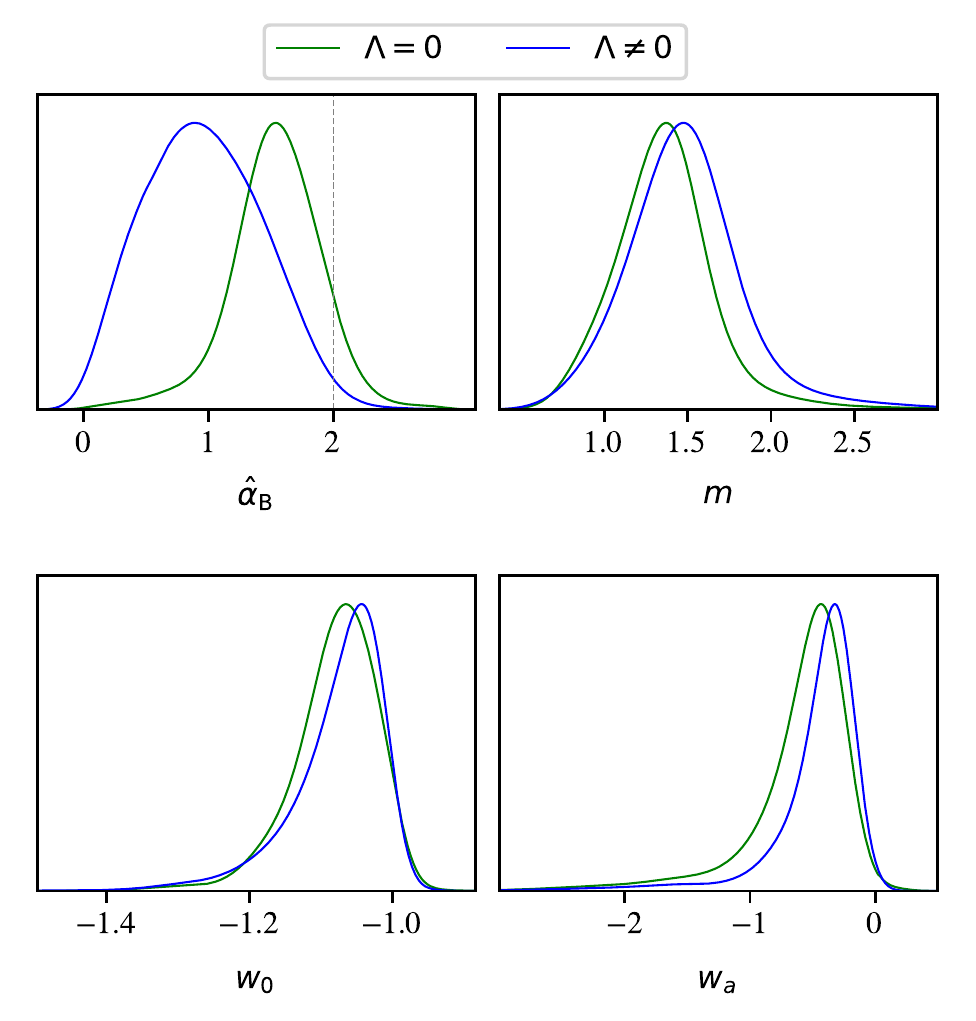}
   \caption{Probability distributions of the $\{w_0,w_a,{\hat \alpha}_{\rm
   B},m\}$ parameters obtained minimising Eq.~\eqref{eq:chi2} as explained in
 Sections~\ref{approx} and \ref{results}. The presence of a cosmological constant term in the theory modifies the probability distributions of the parameters. Their correlations can be seen in Fig.~\ref{fig:aB-w_distrib_fit}. $\Lambda=0$ is the case where $\Omega_{\rm DE} = \Omega_\phi$ and $\Lambda\neq0$ is where $\Omega_{\rm DE} = \Omega_\phi + \Omega_\Lambda$. Fig.~\ref{fig:ci_normal-vs-self-accel} shows the different distributions of the $\{c_{01},c_{02},d_{02}\}$ parameters given these two cases. For details on the vertical line at $\hat{\alpha}_{\rm B} = 2$ we refer the reader to the Appendix~\ref{app-disco}.}
  \label{fig:aB-w_distrib}
\end{figure}

\begin{figure*}[t!]
  \centering
   \includegraphics[trim={2cm 2cm 0cm 2cm},clip,width=0.95\textwidth] {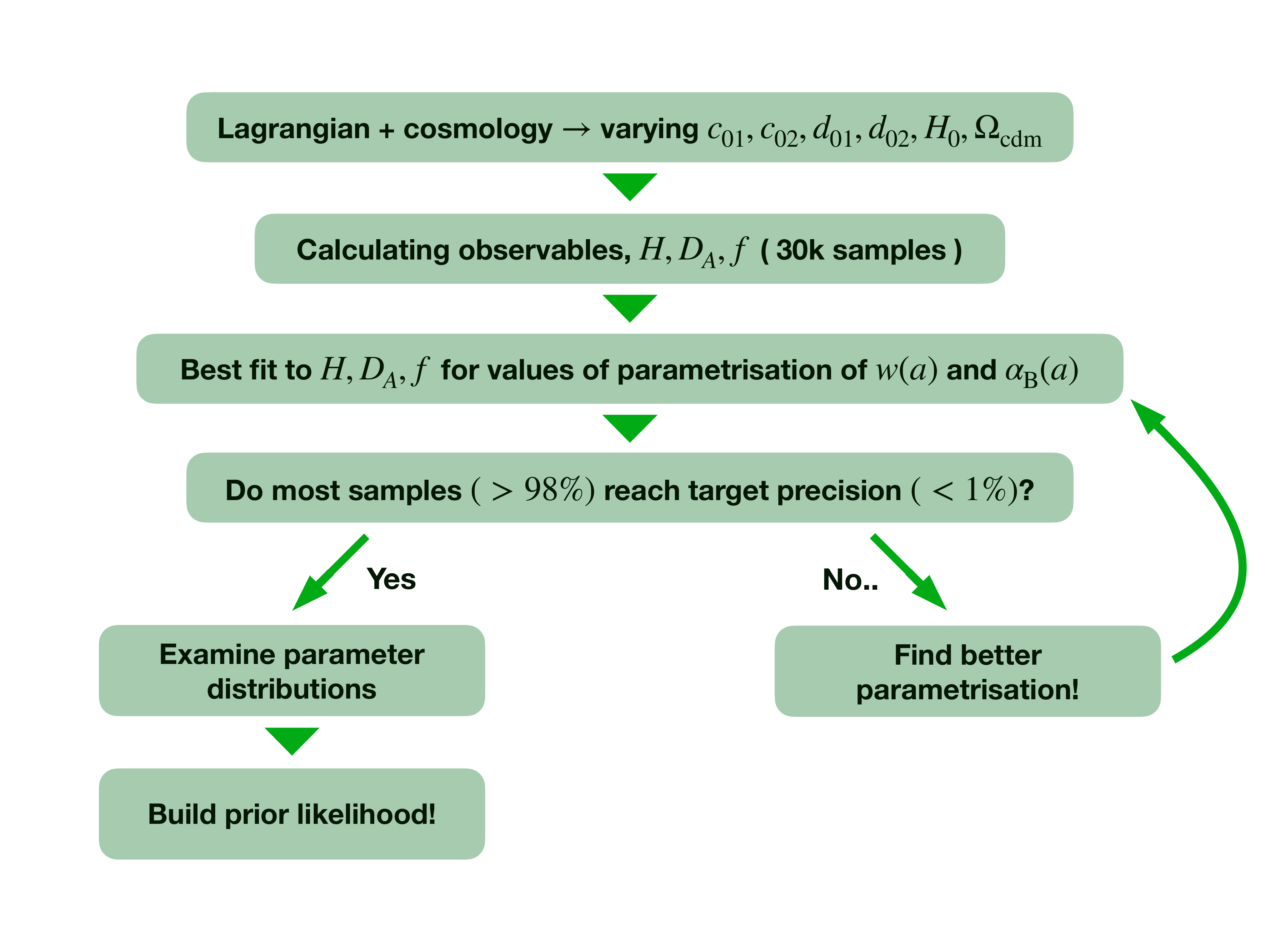}
   \caption{Diagram summarising the method presented in Section~\ref{approx} that we use for determining the correct evolution functions of the model, $w(a)$ and $\alpha_{\rm B}(a)$ and the set of coefficients $\{w_0,w_a,{\hat \alpha}_{\rm B},m\}$ that best fit the observables computed from the Lagrangian.}
  \label{fig:diagram}
\end{figure*}

In the spirit of~\cite{Garcia-Garcia:2019cvr}, we now want to find the set of parameters $\{w_0,w_a,{\hat \alpha}_{\rm B},m\}$ that reproduce the Hubble rate ($H$), the angular diameter distance ($D_A$) and the growth factor ($f=d\ln\delta_m/d\ln a$),
\begin{align}
  H^2 &=\frac{1}{3M_{\rm P}^2}  \sum_i \rho_i\,,\\
  D_A &= \int_0^z \frac{dz'}{H(z)}\,,\\
f^\prime &+f^2 + \left(2 + \frac{\dot{H}}{H^2}\right)f -
\frac{3}{2}\Omega_m \left(1 + \frac{\alpha_{\rm B}^2}{2 {c_{\rm sN}}^2} \right)
= 0\,,
\label{eq:f}
\end{align}
where
\[
{c_{\rm sN}}^2 =\left[(\alpha_{\rm B}-2)\left(\dot{H}-H^2\alpha_{\rm B}/2\right) + H \dot{\alpha}_{\rm B} - \rho_{\rm m} - p_{\rm m}\right]/H^2\,,
\]
computed from the Lagrangian with parameters $(c_{01}, c_{02}, d_{01}, d_{02})$ with the accuracy required by next-generation surveys; i.e. $1\%$ at $z < 10$~\cite{Abell:2009aa,Font-Ribera:2013rwa,Aghamousa:2016zmz,Joudaki:2017zhq}. and $0.3\%$ at recombination for $D_A$~\cite{Aghanim:2018eyx}, for  99\% of the models.
In order to find $\{w_0,w_a,{\hat \alpha}_{\rm B},m\}$ we minimise
\begin{equation}
  \chi^2 = \sum_z\frac{(\mathcal{O}(w_0,w_a,{\hat \alpha}_{\rm B},m)_z - \mathcal{O}(c_{01}, c_{02}, d_{01}, d_{02})_z)^2}{\sigma_{\mathcal{O}_z}^2}\,,
  \label{eq:chi2}
\end{equation}
where $\mathcal{O}(w_0,w_a,{\hat \alpha}_{\rm B},m)_z$ and $\mathcal{O}(c_{01}, c_{02}, d_{01}, d_{02})_z$ are the observables at redshift $z$ computed with the parametrisations of $w$ and $\alpha_{\rm B}$ and the exact from the full evolution of the field equations for a shift-symmetric model given by the set of parameters $\{c_{01}, c_{02}, d_{01}, d_{02}\}$, respectively.
The variable $\sigma_{\mathcal{O}_z}$ weights each point so that we can require different precision depending on the variable and redshift.
For instance, we set $\sigma_{\mathcal{O}_z} = 10^{-3}$  for all observables at $z<10$ and $\sigma_{D_{\rm A}(z_{\rm rec})} = 10^{-4}$ at recombination for the angular diameter distance. 
The software used to make these fits is a modified version of RUFIAN~\cite{Garcia-Garcia:2019cvr} and can be found at \url{https://gitlab.com/dinatraykova/horndeski-priors}. 

Let us emphasise that, with this approach, we do not choose the set $\{w_0, w_a, \hat\alpha_{\rm B}, m\}$ that best fit the equation of state and $\alpha_{\rm B}$ curves obtained from the Lagrangian, which are not observable quantities.
Instead, we minimise the error in the background evolution $H$ and $D_A$, and $f$ for the linear perturbations.
In this sense, allowing  $\{w_0, w_a, \hat\alpha_{\rm B}, m\}$ to differ from their best fit values with respect to the exact $w$ and $\alpha_{\rm B}$, we find the set of parameters that minimise the error on the observables.
It is important to note that, in comparison with quintessence, the equation of $f$, Eq.~\eqref{eq:f}, has the source term modified as
   \begin{equation}
     \frac{3}{2} \Omega_{\rm m} \longrightarrow \frac{3}{2}\Omega_m \left(1 + \frac{\alpha_{\rm B}^2}{2 {c_{\rm sN}}^2} \right) \,,
   \end{equation}
which introduces an extra dependency on $\alpha_{\rm B}$ in the source term and means that a more precise fit to $\alpha_{\rm B}$ would be required to get a good fit to the observables than is the case for $w$.

In Fig.~\ref{fig:aB-w_distrib} we show the distributions for $\{w_0, w_a, \hat\alpha_{\rm B}, m\}$ set of parameters for the shift-symmetric model with and without $\Lambda$ recovered by minimising the error on the observables, as detailed here; correlations between these variables will become apparent as we construct a complete model for the priors in the next section. 

In Fig.~\ref{fig:diagram}, we present a summary diagram of the method explained in this Section that allows us to derive the approximate time dependent functions that describe the shift-symmetric model, $w(a)$ and $\alpha_{\rm B}(a)$, and find the best fit coefficients $\{w_0,w_a,{\hat \alpha}_{\rm B},m\}$ that better reproduce the observable quantities obtained from the evolution of the Lagrangian $\{c_{0i},d_{0i}\}$.

\section{Results}
\label{results}
\begin{figure}[t!]
  \centering
  \includegraphics[width=0.5\textwidth]{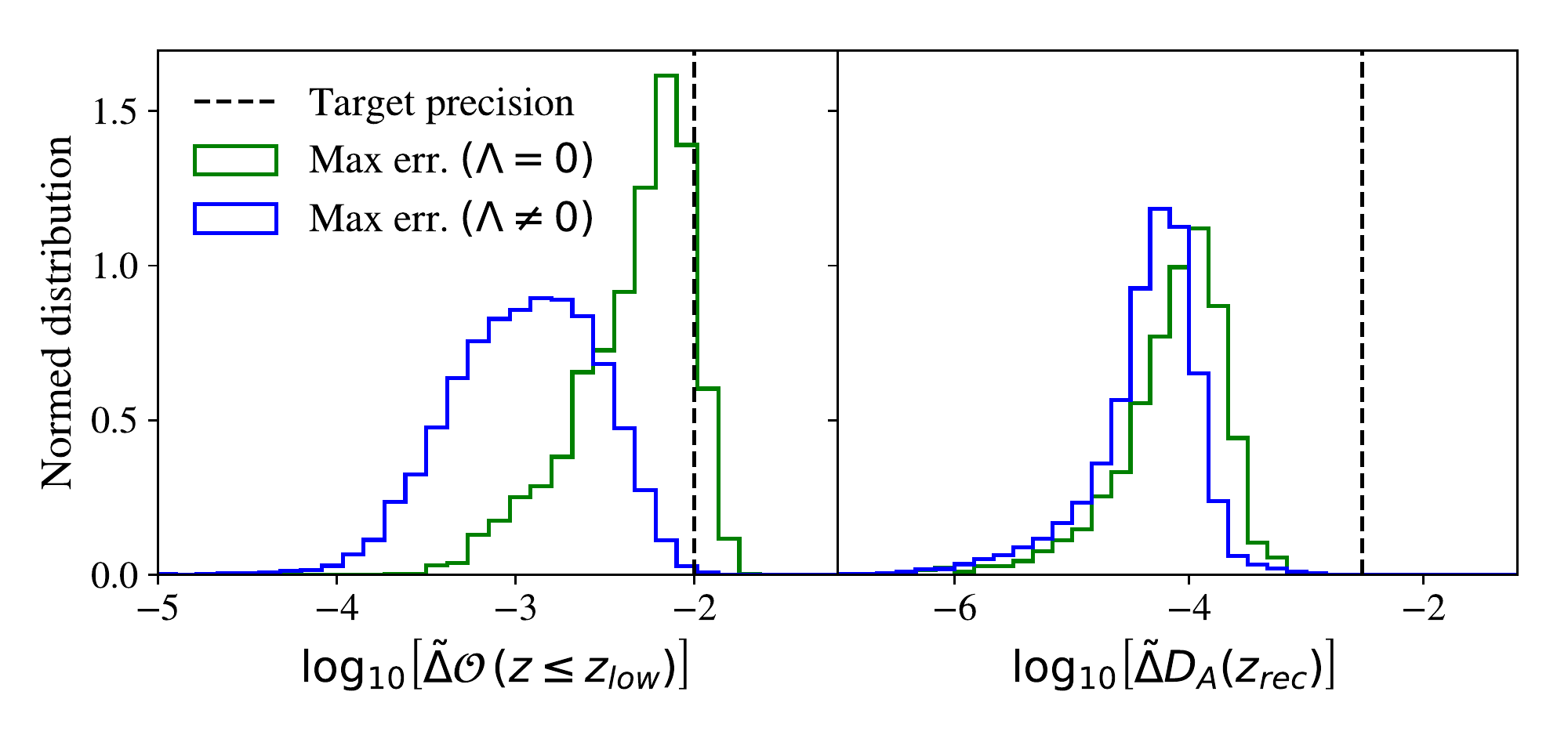}
   \caption{Left panel: distribution of the maximum relative deviation between
     the observables computed from the theory ($\mathcal{O}(c_{01}, c_{02},
         d_{01}, d_{02})$) and the parametrisation ($\mathcal{O}(w_0,w_a,{\hat
         \alpha}_{\rm B},m)$) at $z<10$ for each fit. Right panel:
         distribution of the relative error between the $D_A(z_{\rm rec})$
         computed from the theory and the parametrisation for
     each fit.}
  \label{fig:residuals}
\end{figure}
We now have a robust process for determining $\{w_0,w_a,{\hat \alpha}_{\rm B},m\}$ for each choice of the physical priors:
minimising Eq.~\eqref{eq:chi2} allows us to find the set of parameters  that reproduce the observables $H$, $D_A$ and $f$ with the accuracy needed by next-generation surveys.
The next step is to obtain the {\it probability distribution} that will be used as theoretical priors for the shift-symmetric Horndeski models.
For that, we sample 30,000 random models with parameters $\{c_{01}, c_{02}, d_{01}, d_{02}\}$ and store their corresponding observables at specific redshifts (100 points at $z < 10$ and at $z_{\rm rec}$, in the case of $D_A$).
After minimising Eq.~\eqref{eq:chi2} for each of this set of 30,000 observables, we end up with having 30,000 $\{w_0,w_a,{\hat \alpha}_{\rm B},m\}$ that can be used to build our theoretical priors. 
We obtain the observables quantities for each realisation using \texttt{hi\_class} \cite{Zumalacarregui:2016pph,Bellini:2019syt}, an extension to the Boltzmann code \texttt{CLASS} \cite{Blas:2011rf} that solves the cosmological equations for a broad range of sub-sets of the Horndeski class of theories.

\begin{figure}[t!]
  \centering
   \includegraphics[width=0.48\textwidth]{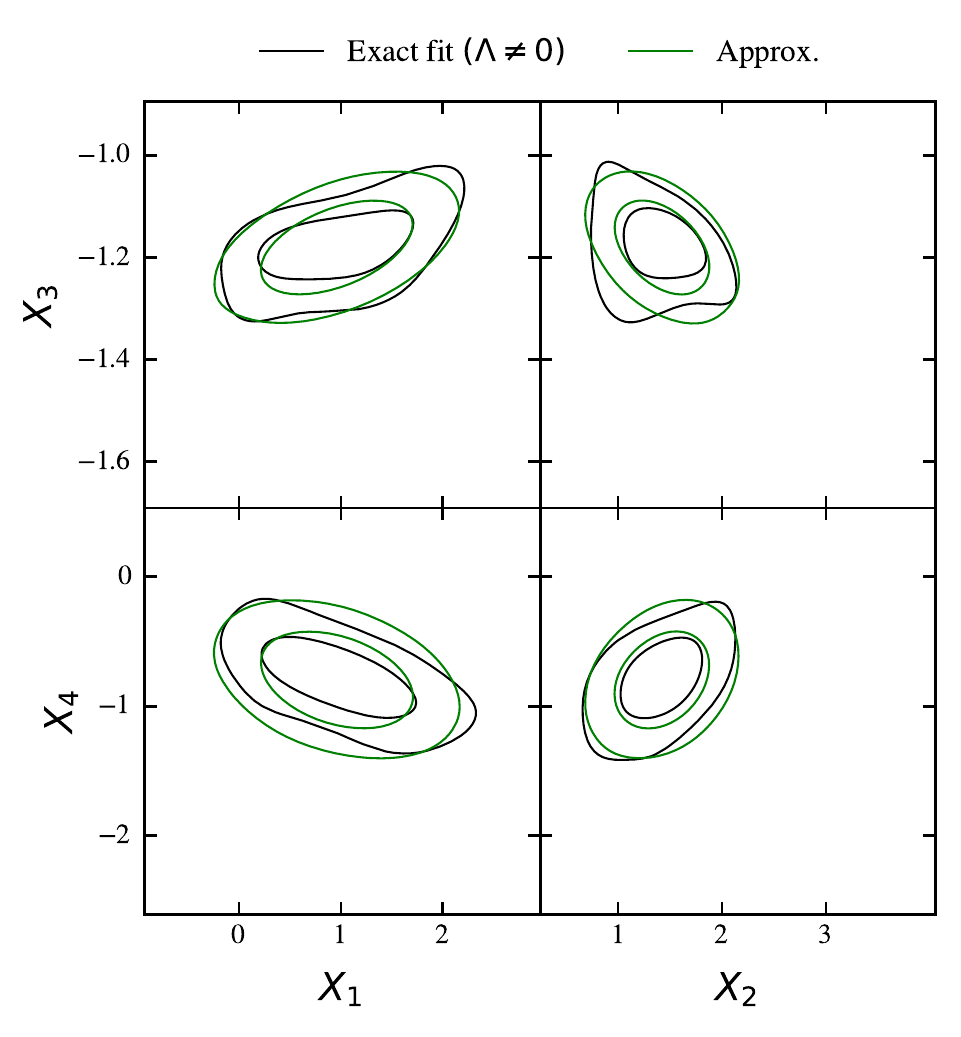}
   \caption{Approximate fit to the probability distribution of the $\{X_1,X_2,X_3,X_4\}$ for the $\Lambda=0$ model. In black, the contours from the original set of parameters transformed by Eq.~\eqref{eq:X-basis}. In green, those obtained from a multivariate Gaussian distribution. As one can see, their differences are small and have little effect on the original parameters $\{w_0,w_a,{\hat \alpha}_{\rm B},m\}$ (Fig.~\ref{fig:aB-w_distrib_fit}), the observables (Fig.~\ref{fig:observables}) and, therefore, in an MCMC.}
  \label{fig:Xi_distrib}
\end{figure}
\begin{figure*}[htb!]
  \centering
   \includegraphics[width=0.95\textwidth]{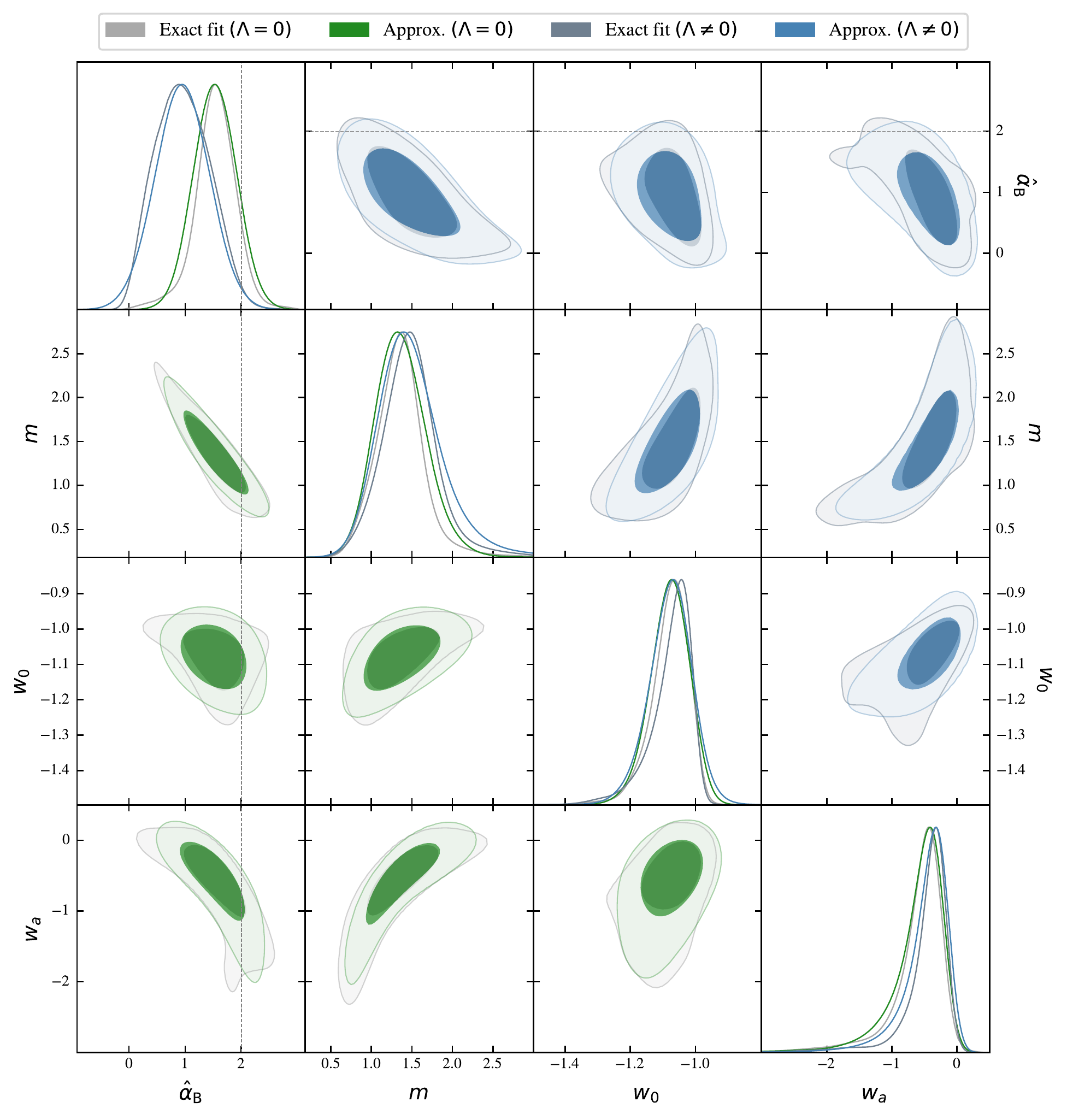}
\caption{Probability density distributions of the $\{w_0,w_a,{\hat \alpha}_{\rm B},m\}$ parameters for the $\Lambda=0$ (lower left) and $\Lambda\neq0$ (upper right) shift-symmetric theories studied. We compare the exact distributions obtained fitting the observables with those obtained sampling from the new Gaussianised space $\{X_1,\, X_2,\, X_3\, X_4\}$ and transforming back using Eq.~\eqref{eq:X-basis}. The differences are small and do not affect the observables significantly (Fig.~\ref{fig:observables}). Here we have highlighted the $\alpha_{\rm B} = 2$ line to separate the region where the evolutions can display some singular behaviour (see Appendix~\ref{app-disco}).}
  \label{fig:aB-w_distrib_fit}
\end{figure*}
\begin{figure*}[htb]
  \centering
   \includegraphics[width=0.95\textwidth]{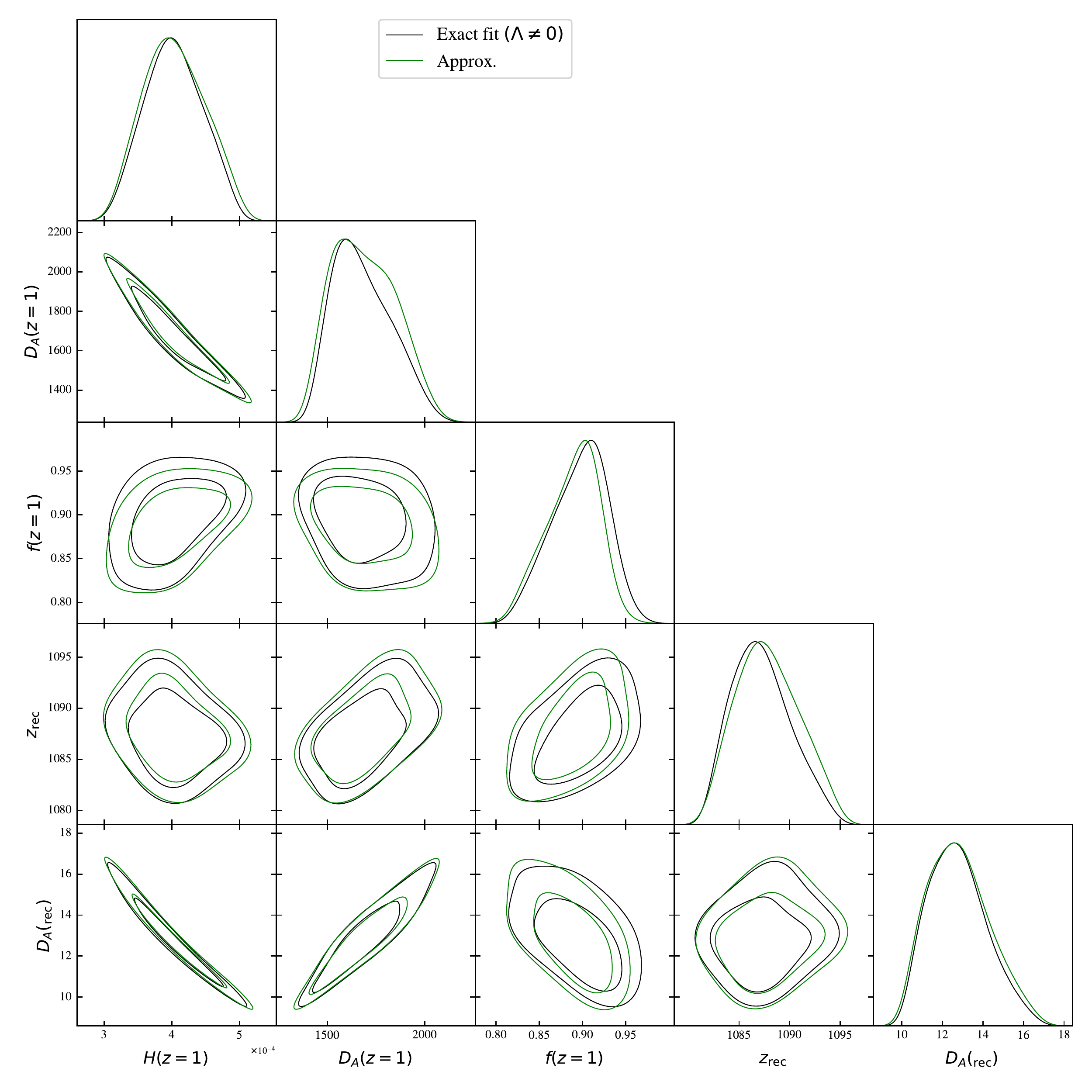}
   \caption{Distributions of the observables at $z=1$ and recombination obtained integrating the field equations of motion (black lines) compared to those obtained using the parameters $\{w_0,w_a,{\hat \alpha}_{\rm B},m\}$, recovered after sampling from the multivariate Gaussian distribution (green lines) in the $\{X_1, X_2, X_3, X_4\}$ space, with Eq.~\eqref{eq:X-basis}. As we advanced, the differences on the observables are small, which justifies the Gaussian approximation we used and gives confidence to the theoretical priors we have built, Eqs.~\eqref{eq:pdf}--\eqref{eq:mu_sigma_Lneq0}).}
  \label{fig:observables}
\end{figure*}

Let us note that the choice of 30,000 samples and 100 points at $z<10$ is just a matter of computational efficiency and has no physical insights.
We checked that after 30,000 samples the probability distributions had already converged and increasing its size to 100,000 does not alter the results.
In addition, for each set of $\{w_0,w_a,{\hat \alpha}_{\rm B},m\}$ obtained minimising Eq.~\eqref{eq:chi2} with 100 points for each observable below $z=10$ and $D_A$ at $z_{\rm rec}$, we saw that the new observables satisfy the requirement of having an error below $1\%$ at $z < 10$ and below $0.3\%$ for $D_A(z_{\rm
rec})$ for the $99\%$ of the cases.
This can be seen in Fig.~\ref{fig:residuals}. 

The probability distributions for the parametrised shift-symmetric Hordenski models can be seen in Fig.~\ref{fig:aB-w_distrib},
which shows mild correlations between different parameters. 
We note that these correlations do not have the usual elliptical shape that one expects for a multivariate Gaussian. 
Clearly there is a non-linear correction that must be taken into account in the next steps.

If we are to construct theoretical priors that can be used in a
Markov chain Monte Carlo (MCMC), we need to find a sufficiently good approximation of the probability distribution that allows us to recover the same distribution of the parameters and the observables when sampling from it. 
We do this in two steps. We first transform to a new set of parameters
\begin{eqnarray}
X_1&=& \hat{\alpha}_{\rm B}\,, \nonumber \\
X_2&=&m\,\hat{\alpha}_{\rm B}^{1/6}\,, \nonumber \\
X_3 &=&w_0\,m^{1/4}\,, \nonumber \\
X_4&=& w_a\,m^2\,,
\label{eq:X-basis}
\end{eqnarray}
which effectively Gaussianise the distributions (Fig.~\ref{fig:Xi_distrib}). 

We find that a multivariate normal distribution fits the distribution of $\{X_1,X_2,X_3,X_4\}$ and that, once transformed back to $\{w_0,w_a,{\hat \alpha}_{\rm B},m\}$ recovers the correlations between the variables to a very good approximation; this can be seen in Fig.~\ref{fig:aB-w_distrib_fit}.
Of course, a crucial test is to see the impact on the observables and whether one is able to recover the correct distribution for those.
We do so in Fig.~\ref{fig:observables} where we compare the observables obtained by integrating the field equations of motion to the ones obtained by sampling from the distribution. 
Note that we only show the distances at $z=1$ and at recombination, as they have the largest differences, yet these are still small and should have little impact on the posterior distributions of the parameters when combined with data.

Using the Gaussianised distribution we can construct an analytic model and calculate the probability density in the transformed parameter basis, 
\begin{equation}
    p(X|\mu,\Sigma) = \frac{1}{(2\pi)^2|\Sigma|^{1/2}}\exp\left(-\frac{1}{2}(X-\mu)^T\,\Sigma^{-1}\,(X-\mu)\right)\,
    \label{eq:pdf}
\end{equation}
where $\mu$ is the vector of mean values and $\Sigma$ is the covariance matrix of our prior parameter distribution in the transformed basis $\{X_1,X_2,X_3,X_4\}$. For the model without $\Lambda$ we find

\begin{equation}
\begin{aligned}
    \mu(\Lambda=0) &= \left(1.5346, 1.4461, -1.1592, -0.8841\right)\,,\\
\Sigma (\Lambda=0) &= 
\begin{pmatrix}
  0.1475 & -0.0916 &  0.0160 & -0.0469 \\
 -0.0916 &  0.0776 & -0.0087 &  0.0326 \\
  0.0160 & -0.0087 &  0.0041 & -0.0079 \\
 -0.0469 &  0.0326 & -0.0079 &  0.0516 \\
\end{pmatrix}\,.
\end{aligned}
\label{eq:mu_sigma_L0}
\end{equation}
For the case with $\Lambda$ we have
\begin{equation}
\begin{aligned}
    \mu(\Lambda\neq0) &= \left(0.9545, 1.4255, -1.1803, -0.7962\right)\,,\\
\Sigma (\Lambda\neq0) &= 
\begin{pmatrix}
  0.2311 & -0.0511 &  0.0136 & -0.0415 \\
 -0.0511 &  0.0875 & -0.0075 &  0.0218 \\
  0.0136 & -0.0075 &  0.0035 & -0.0017 \\
 -0.0415 &  0.0218 & -0.0017 &  0.0589 \\
\end{pmatrix}\,.
\end{aligned}\label{eq:mu_sigma_Lneq0}
\end{equation}

We use this to infer the a priori likelihood of a given sample used in the combined analysis with data in Section~\ref{data}.

We might want to compare our current results with those of our previous work~\cite{Garcia-Garcia:2019cvr}. 
There, we found that the phenomenology of different highly dimensional theories is well described by the usual $w_0$-$w_a$ parametrisation. 
In that case, we went from having many parameters in the Lagrangian to just two, accurately accounting for their cosmologies. 
In this paper, however, we start from a Lagrangian with three parameters (after fixing $d_{01} = -1$) and end up with four parameters to describe its phenomenology. However, this phenomenological parametrisation is still advantageous: it does not require solving the field equations, allows to clearly split the background and linear perturbations effects and shows that both $w$ and $\alpha_{\rm B}$ are simpler than one would a priori think. We expect this to also be the case in other more general theories.

\section{Comparison with current data}
\label{data}
\begin{figure*}[t!]
  \centering
   \includegraphics[width=0.95\textwidth]{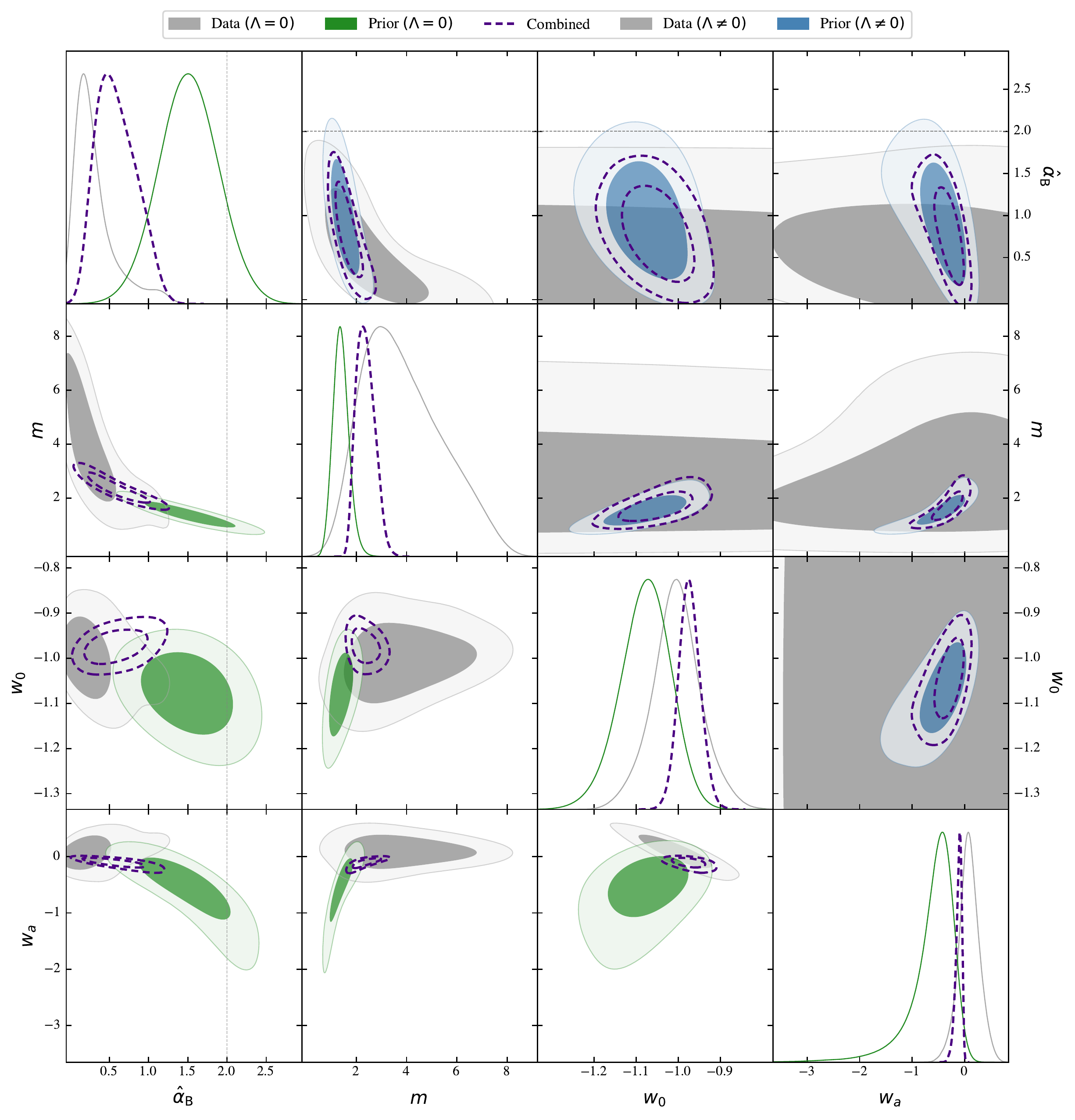}
   \caption{Comparison between data and the prior distributions for the two shift-symmetric variants we have considered, $\Lambda=0$ (lower left corner) and $\Lambda\neq0$ (upper right). 
   The grey filled contours show the distributions of the parameters when constrained with `data' alone, the green or blue filled contours show our `priors' for the $\Lambda=0$ and $\Lambda\neq0$ cases respectively, and the purple dashed contours are from the combined analysis.
   Note that the `prior' likelihood is built in the Gaussianised $\{X_i\}$ basis, as discussed in Sections~\ref{results} and \ref{data} and takes into account the underlying physical properties of the model (not to be confused with the flat uncorrelated priors we put on the parameters in the `data' runs).
   We have only plotted the 1D pdf's for the case $\Lambda=0$ for clarity and again we have marked the line where $\alpha_{\rm B} = 2$ (see Appendix~\ref{app-disco}). 
   This plot demonstrates how combining data with theoretical priors can result in much tighter constraints on some of the parameters of the model.}
  \label{fig:priors-vs-data}
\end{figure*}
Here we present the constraints on $\alpha_{\rm B}$ and $w$ from current cosmological data and compare and combine these with our theoretical priors.
To do this we use a combination of cosmic microwave background (CMB), Baryon Acoustic Oscilations (BAO), Redshift Space Distorsions (RSD) and Supernovae Type IA (SN Ia) data.

From Planck 2018 \cite{Akrami:2018vks,Aghanim:2019ame,Aghanim:2018eyx}, we use the auto- and cross-correlations of the temperature ($T$) and polarisation ($E$) fluctuations of the Cosmic Microwave Background, together with measurements of the lensing potential; i.e. the likelihood from the high-$l$ temperature auto-correlation (TT), temperature  and polarisation cross-correlation (TE), and the polarisation auto-correlation (EE) spectra (at $l\geq30$), the low-$l$ ($2\leq l < 30$)) TT and EE likelihoods and the lensing likelihood (with temperature and polarisation lensing reconstruction) in the multipole range $l=8-400$.

Additionally we use the BAO and RSD measurements from BOSS DR12 \cite{Alam:2016hwk}, as well as BAO from the 6dFGS survey \cite{Beutler:2011hx}.
The BAO measurements are of the Hubble rate, $H$ and the angular diameter distance $D_A$, while RSD measures the growth rate of the universe through $f(z)\sigma_8(z)$.
We use the full covariance between the $f(z)\sigma_8(z)$ measurements at different redshifts and the BAO measurements of $H(z)$ and $D_{\rm A}(z)$ from BOSS.
However, we do not consider the correlation between the BOSS and 6dF measurements as those cover different areas of the sky and thus any such correlation would be negligible.

Finally, we also include the Pantheon SNe Ia sample \cite{Scolnic:2017caz}, which 
combines the Pan-STARRS1 Medium Deep Survey with ones from the SDSS, SNLS, and various low-redshift and HST samples, 1048 SNe Ia in total in the redshift range $0.01 < z < 2.3$.
We also note that, throughout, we assume that the cross-correlation between the different datasets is negligible.

We built our prior likelihood in the Gaussianised basis $\{X_1,X_2,X_3,X_4\}$ and, in order to be consistent, we use the same basis for the sampling in all cases, from which we then convert the resulting distributions back to $\{w_0,w_a,{\hat \alpha}_{\rm B},m\}$. 
We sample through the parameter set $\{X_1,X_2,X_3,X_4\}$
together with the standard cosmological parameters in a Markov Chain Monte Carlo (MCMC) with \texttt{MontePython} \cite{Audren:2012wb,Brinckmann:2018cvx} using
the Metropolis-Hastings algorithm \cite{Metropolis:1953am,Hastings:1970aa}. 
We do not consider any prior bounds on the standard cosmological parameters in the MCMC run (apart from $\tau > 0.004$), but just start from a known good fit point from Planck for the set $\{\Omega_{\rm cdm, 0}, \Omega_{\rm b, 0}, H_0, A_s, n_s, \tau\}$.
For the $X_i$ we set the following ranges,
\begin{align}
    &X_1 \in (0, 10)\,,\qquad X_3 \in (-10, 0)\,,\\
    &X_2 \in (0, 15)\,,\qquad X_4 \in (-15,30)\,.
\end{align}
In the case of $\Omega_\Lambda\neq0$ we also set $\Omega_\Lambda\in (0,1)$.
Using the Gelman-Rubin convergence criterion \cite{Gelman:1992zz} we require $R-1<0.02$. The contour plots were produced using GetDist \cite{Lewis:2019xzd}.

In order to obtain the combined constraints from data and the theoretical priors, we implemented Eqs.~\eqref{eq:pdf}--\eqref{eq:mu_sigma_Lneq0} in \texttt{MontePython} as a new likelihood module\footnote{Available at:  \url{https://github.com/dinatraykova/shift_priors}}. 
In the analysis with data only we assume uniform priors on the $\{X_1,X_2,X_3,X_4\}$ parameters.

We present the results in Fig.~\ref{fig:priors-vs-data}, where in the bottom half of the triangle we show the contours for the $\Lambda=0$ case of the shift-symmetric model and on the top half we have the case with additional $\Lambda$.
On this plot we show the parameters data constraints (grey solid line, filled contours) overlaid with the distributions of the priors. (in green for $\Lambda=0$ and blue for  $\Lambda\neq0$). 
In addition, we show the combined constraints (dashed line) from both the data and the priors. 
In this figure, note that we only show the $\Lambda=0$ 1D distributions since the $\Lambda\neq0$ data-only constraints are much broader than any of the others, making it difficult to read.

\begin{figure}[t!]
  \centering
  \includegraphics[width=0.4\textwidth]{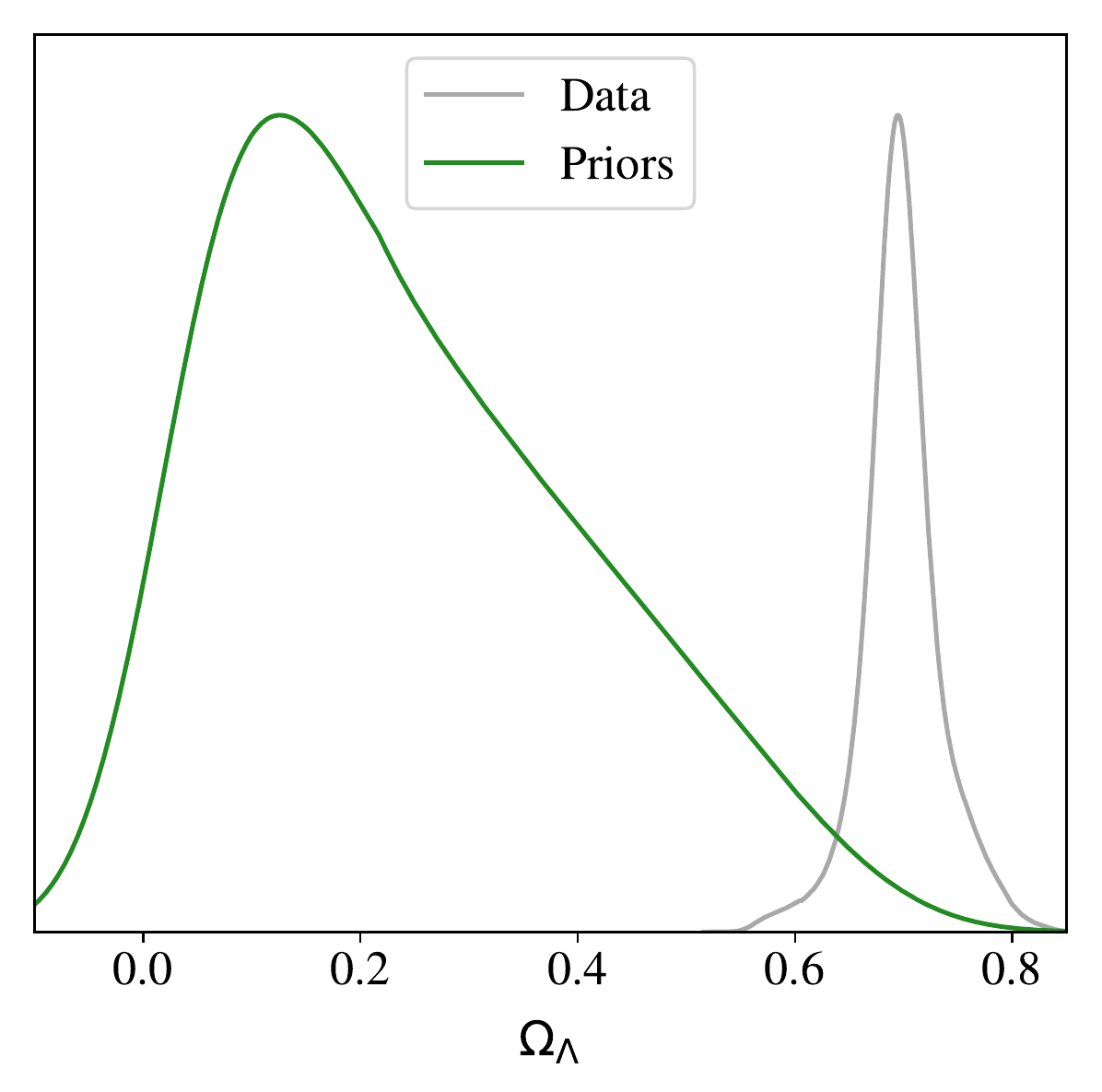}
   \caption{Probability distributions of $\Omega_\Lambda$ from the analysis with data alone (grey solid line), our priors (green solid line). Let us note that $\Omega_\Lambda \sim 0.7$ in the data run implies that data seems to prefer a universe mainly filled with a cosmological constant and only a small contribution of the scalar field to DE.}
  \label{fig:omega_lambda}
\end{figure}

If we focus only on the bottom left corner of Fig.~\ref{fig:priors-vs-data} ($\Lambda=0$), we note that although the data and the priors appear to be equally as constraining for ${\hat \alpha}_{\rm B}$, $w_0$ and $w_a$, the contours are misaligned and only overlap away from their respective centres (i.e. regions of high probability). 
Nevertheless they are statistically consistent and combining the two results in tighter bounds on these parameters.
Further, data alone does not provide a very strong bound on $m$, which makes including the prior likelihood crucial in constraining the time evolution of $\alpha_{\rm B}$.
Looking at the combined constraints (dashed line) one may expect that the contours should lie in between the data and priors alone. 
However, due to the high dimensionality of the problem and the correlations between different parameters, the 2D and 1D projections of the distributions from the combined analysis can end up off the centre of the distributions of the data and the priors runs alone. 
In the case of $w_0$ this effect can be seen quite clearly, where the combined histogram appears to be to the right of both the data and priors alone. 
This is not surprising looking at the 2D contours of $w_0$ and the other three parameters, where we see that the data and priors contours overlap only at their edges, which could result in such a shift in the 1D projections of one or more parameter.

In the top right corner of Fig.~\ref{fig:priors-vs-data} we show the contours for the case where we include a contribution of $\Lambda$ to the DE density ($\Lambda\neq0$).
In this case we find that while data alone (grey solid contours) constrains the $\hat{\alpha}_{\rm B}$ and $m$ parameters well, the distributions of $w_0$ and $w_a$ are very wide, compared to the priors (blue solid contours) and the combined constraints are almost fully driven by our priors.
We note that these distributions cover almost the full range that we have set for these parameters in the data run.
The distributions of $\Omega_\Lambda$ are shown in Fig.~\ref{fig:omega_lambda} from the sampling with data (grey line) and the ones recovered when deriving the priors (green line). 
This plot shows that the data prefers the majority of the DE density contribution to come from $\Lambda$, as it is consistent with $\Omega_\Lambda \sim 0.7$.
This leaves only a very small portion of $\Omega_{\rm DE}$ to come from the scalar field $\phi$.
Our definition of $w(a)$, Eq.~\eqref{eq:rho_p} includes only the contribution of the scalar field, so we can expect that in a case where $\Omega_\Lambda$ dominates the DE density, it would not be possible to get a strong constraint on the equation of state parameters of the field, $w_0$ and $w_a$.

\begin{table}[t!]
\begin{center}
\begin{tabular}{|p{1.5cm}||p{1.3cm}|p{1.3cm}|p{1.8cm}|p{1.75cm}|}
 \hline
     & $\hat{\alpha}_{\rm B}$ & $m$ & $w_0$ & $w_a$\\
 \hline
 ($\Lambda=0$) & & & & \\
 \hline
 Data       & $0.3 \pm 0.3$
            & $3.8 \pm 1.6$
            & $\hspace{0.15cm}-1.0 \pm 0.06$
            & $\hspace{0.4cm}0.1 \pm 0.2$\\
 \hline
 Prior      & $1.5 \pm 0.4$
            & $1.4 \pm 0.4$
            & $-1.08 \pm 0.06$
            & $\hspace{0.15cm}-0.6 \pm 0.4$\\
 \hline
 Combined   & $0.6 \pm 0.3$
            & $2.4 \pm 0.4$
            & $-0.97 \pm 0.03$
            & $-0.11 \pm 0.06$\\
  \hline
  \hline
    ($\Lambda\neq0$) & & & &\\
  \hline
 Data       & $0.5 \pm 0.4$
            & $2.6 \pm 1.4$
            & $\hspace{0.7cm}--$
            & $\hspace{0.55cm}--$\\
 \hline
 Prior      & $1.0 \pm 0.5$
            & $1.5 \pm 0.4$
            & $-1.08 \pm 0.07$
            & $-0.5 \pm 0.4$\\
 \hline
 Combined   & $0.8 \pm 0.4$
            & $1.7 \pm 0.4$
            & $-1.05 \pm 0.06$
            & $-0.3 \pm 0.2$\\
 \hline
\end{tabular}
\caption{Best fit and confidence limits of $w_0$ and $w_a$ for the data set CMB+BAO+RSD+SN, the theoretical priors and the combined analysis for the shift-symmetric models both with and without $\Lambda$, ($\Lambda=0$ and $\Lambda\neq0$).
Note that we have not written the means and errors for $w_0$ and $w_a$ from the data run in the $\Lambda\neq0$ case, as data is not constraining on these; the errors are determined by the ranges we have set and the mean values are consistent with $\Lambda$CDM. This is related to the fact that, in this work, $w$ is defined as the scalar field equation of state and that data seems to prefer a negligible contribution of the scalar field to the DE density with the majority coming from $\Lambda$ (Fig.~\ref{fig:omega_lambda}).}
\label{tab:best-fit}
\end{center}
\end{table}

To emphasise the benefit of including theoretical priors into the likelihood analysis in constraining these models, in Table~\ref{tab:best-fit} we present the parameter ranges for the $\{w_0,w_a,{\hat \alpha}_{\rm B},m\}$ set from the likelihood analysis with uniform priors and with theoretical priors. 
In the $\Lambda=0$ case we find that for $m$ there is a significant improvement in the error after including our theoretical priors, compared to the constraints with data using uniform uncorrelated priors (from $\pm1.6$ to $\pm0.4$). 
For the other three parameters ranges from the data run with uncorrelated priors and our derived correlated ones are comparable but combining them still results in slight improvement of the errors.
In the case of $\Lambda\neq 0$, we see that there is a similar improvement on the constraint of $m$ as we find in the $\Lambda=0$ case (from $\pm1.4$ to $\pm0.4$).
However, as we saw from the contours in Fig.~\ref{fig:priors-vs-data}, $w_0$ and $w_a$ cannot be constrained with data using the uniform priors on the parameter set due data preferring $\Omega_\Lambda$ to be the dominant contribution to the DE density.
The addition of the theoretical priors in this case is, therefore, crucial as the only way to fully constrain the parameter space.

\section{Discussion}
\label{disc}
In this paper we have taken a further step towards constructing a set of physical priors for Horndeski theories of gravity. Building on the experience of constructing such a prior for thawing quintessence, we have focussed on a physically well motivated subset of Horndeski gravity: shift-symmetric theories with standard speed of gravitational waves. While these theories are less general than the full Horndeski space of theories, they are more general than the much studied Galileon scalar-tensor theories.

Working with shift-symmetric theories has allowed us to explore a situation in which one needs more than just the equation of state, $w$, to fully characterise its behaviour on cosmological scales. For such theories one needs to also include an accurate model for the ``braiding'' parameter, $\alpha_{\rm B}$. We have done so, constructing a prior distribution function, ${\cal P}$ for four constant parameters defined in Eqs.~\eqref{eq:w0wa} and \eqref{eq:aB-param}. Remarkably, and very much like in the case of thawing quintessence, we have come up with a simple analytical form for ${\cal P}$ which can be easily deployed in future cosmological parameter analysis. 

We have learnt a number of lessons from focusing on shift-symmetric theories which give us a sense of the challenge of tackling more general Horndeski theories. For a start, the theories we have looked at here are endowed with a tracking behaviour which eliminates the need to pin down a prior for initial conditions. This will not be true in general for full Horndeski theories. 

We have had to face the problem of sampling over a multi-dimensional space of parameters (in this case $\{c_{01},c_{02},d_{01},d_{02}\}$) which is subjected to some form of constraint. The way one implements the constraint can greatly affect the prior distribution function. For example, explicitly solving the constraint can bias the resulting prior, depending on which of the parameters one is solving for. We have argued that one should sample over all parameters and exclude points which lie outside the constraint sub-region. This is, nevertheless, a computationally costly approach to the problem which will become far more severe, the more general the theory one is looking at. 

With an appropriate algorithm for sampling over $\{c_{01},c_{02},d_{01},d_{02}\}$, we have proposed a functional form for the phenomenological parameters, $w$ and $\alpha_{\rm B}$. We have found that the usual form for $w$ is still remarkably effective while, building on our knowledge of Galileons, we have come up with a suitably simple form for $\alpha_{\rm B}$, if we choose the parameters by minimising the error on the observables (a crucial aspect of this approach). The latter insight is useful and points to the fact that, in general, the $\alpha_X$ parameters may have a simple functional form in the more general theory. This means that a reanalysis of current cosmological data may lead to far tighter constraints than have until now been found. 

An important step has been to find non-linear transformations that, to some extent, ``Gaussianises'' the distributions of our parameters. Such a transformation has been remarkably effective allowing us to determine, rather more easily than one would naively expect, an analytic expression for the prior. Again, one would expect this approach to be useful when looking at more general theories.

An interesting aspect of the theories we have focused on -- shift-symmmetric theories -- is that they are, in some sense, viable and complete. In other words, Including terms $\propto X^2$ in $ G_2,G_3$ give viable generalisations of the cubic Galileon (Fig. \ref{fig:priors-vs-data}), even with $\Lambda=0$. In this model $\dot\phi^2/\Lambda_2^4\lesssim 0.2$ (Fig.~\ref{fig:dotphi}), suggesting that higher order corrections are subdominant and can be neglected.

An important aspect, which from our understanding has been somewhat unexplored, is that $c_{01}< 0$ is more generic than just for the Covariant Galileon. This is important, since this regime disconnects these theories from the Minkowski solution and constraints derived for that solution. Note that there may be other solutions where, for example, higher orders in $X^n$ contribute. In that situation, the constraints on $c_{01}$ may be markedly different.

Note that, while we have considered and taken into account a number of theoretical priors and observational constraints throughout this paper, these are of course not complete and one may wish to add additional priors/constraints to this analysis in the future. One such example to highlight are constraints from dark energy-gravitational wave interactions, specifically related to dark energy (gradient) instabilities that can be induced by gravitational wave sources such as massive binaries \cite{Creminelli:2019kjy}. Requiring the absence of these instabilities in general can be used to significantly tighten cosmological parameter constraints \cite{Noller:2020afd}. In the specific shift-symmetric context of the theories considered here, avoiding such instabilities amounts to requiring $\left|\alpha_{\rm B} \right| \lesssim 10^{-2}$. This effectively renders the cubic Horndeski interactions we have considered into an afterthought for cosmology. We will leave a more detailed investigation of this and other additional priors in the context of shift-symmetric theories for future research.

Finally, our brief comparison with current data shows that this theory is a viable, self-accelerating model of the Universe: the physical priors are consistent with the cosmological constraints. This is somewhat promising given the dearth of theoretically viable models of self-acceleration which are currently compatible with cosmological data. A thorough analysis of shift-symmetric cosmologies, along the lines of what has been undertaken in \cite{Joudaki:2020shz} will allow us to assess if such shift-symmetric gravity is a credible contender for the late time acceleration of the Universe.

\section*{Acknowledgments}
\vspace{-0.2in}

\noindent We thank David Alonso, Harry Desmond, Shahab Joudaki, Eva Muller and Ignacy Sawicki for useful discussions.
 D.T., E.B, P.G.F and C.G.G are supported by European Research Council Grant No:  693024 and the Beecroft Trust.
C.G.G. was also supported by PGC2018-095157-B-I00 from Ministry of Science, Innovation and Universities of Spain and by the Spanish grant, partially funded by the ESF, BES-2016-077038.
JN is supported by an STFC Ernest Rutherford Fellowship, grant reference ST/S004572/1.

\ul{Software:} We made extensive use of {\tt numpy} \citep{oliphant2006guide, van2011numpy}, {\tt scipy} \citep{2020SciPy-NMeth} and  {\tt matplotlib} \citep{Hunter:2007} python packages. In addition, the shift-symmetric models were implemented in \texttt{hi\_class} \citep{Zumalacarregui:2016pph, Bellini:2019syt,Blas:2011rf}, the fits to the observables were done with a modified version of RUFIAN\footnote{Original: \url{https://gitlab.com/carlosggarcia/horndeski-priors}. 
This project: \url{https://gitlab.com/dinatraykova/horndeski-priors}} \cite{Garcia-Garcia:2019cvr}, the MCMC were run with \texttt{MontePython} \cite{Audren:2012wb,Brinckmann:2018cvx} and the contour plots were produced using GetDist \cite{Lewis:2019xzd}.

\appendix

\section{Discontinuities for models crossing \texorpdfstring{$\alpha_{\rm B} = 2$}{Lg}} \label{app-disco}

In the main text we briefly alluded to potential issues associated with crossing $\alpha_{\rm B} = 2$. As shown in Fig.~\ref{fig:aB_w-evol}, $\alpha_{\rm B}$ generically starts strongly suppressed at high redshifts and then grows towards redshift zero, typically reaching ${\cal O}(1)$ values. For a small, yet significant, subset of the models discussed in this paper (see e.g.~the distributions shown in Fig.~\ref{fig:aB-w_distrib}) $\alpha_{\rm B}$ eventually grows to be larger than 2. This is important, because crossing the $\alpha_{\rm B} = 2$ point is associated with a number of discontinuities. This was first noted in \cite{Bellini:2014fua} and discussed in \cite{Lagos:2017hdr,Ijjas:2017pei}. As a result, evolutions crossing this point have being conservatively excluded in some of the subsequent analyses -- see e.g.  \cite{Noller:2018wyv,SpurioMancini:2019rxy,Noller:2020afd}. On the other hand, this could be only a gauge discontinuity, as advocated in \cite{Ijjas:2017pei}, that can be safely removed. So in this appendix we quickly summarise the issues associated with crossing this point and how we treat models that do so in this paper. 

\noindent {\it Discontinuity in the number of propagating degrees of freedom}: Horndeski scalar-tensor models of dark energy generically propagate two scalar degrees of freedom: one directly associated to dark energy and one to matter. Following the approach outlined in \cite{Bellini:2014fua,Lagos:2016wyv,Lagos:2017hdr} and for concreteness modelling matter as a minimally coupled canonical scalar field $\psi_M$ with Lagrangian ${\cal L} = -\tfrac{1}{2}\partial_\mu \psi_M \partial^\mu \psi_M - V(\psi_M)$, working on a cosmological background and in unitary gauge we find these two independent degrees of freedom can be associated with $\delta\psi_M$ and $\Phi$ (i.e.~the scalar metric perturbation of the $ii$ component of the metric). While crossing $\alpha_{\rm B} = 2$ in the evolution, the following constraint relating these two degrees of freedom emerges
\begin{align} \label{eq:aB2constraint}
&\psi_M^{\prime} \delta \psi_M = 2 M_P^2 \Phi^{\prime},
\end{align}
where we have assumed that there is no cosmological running of the Planck mass, as is the case for the models considered in this paper. This relation shows that one propagating degree of freedom is eliminated at this point, so only one dynamical degree of freedom remains here. This is alarming, since the number of propagating degrees of freedom therefore changes as we evolve through $\alpha_{\rm B} = 2$: it is two on either side, but only one remains on the divide itself. 
This may be an artefact of using perturbation theory, but in any case indicates a  potential ill-definedness in the evolution across $\alpha_{\rm B} = 2$. Giving a definitive answer may require a non-perturbative analysis, which allows to follow the dynamics of the real degree of freedom (and not an approximated version of it), and may allow us to exclude the dangerous situation of being in a strongly coupled regime.
\\
\begin{figure}[t!]
  \centering
   \includegraphics[width=0.95\columnwidth]{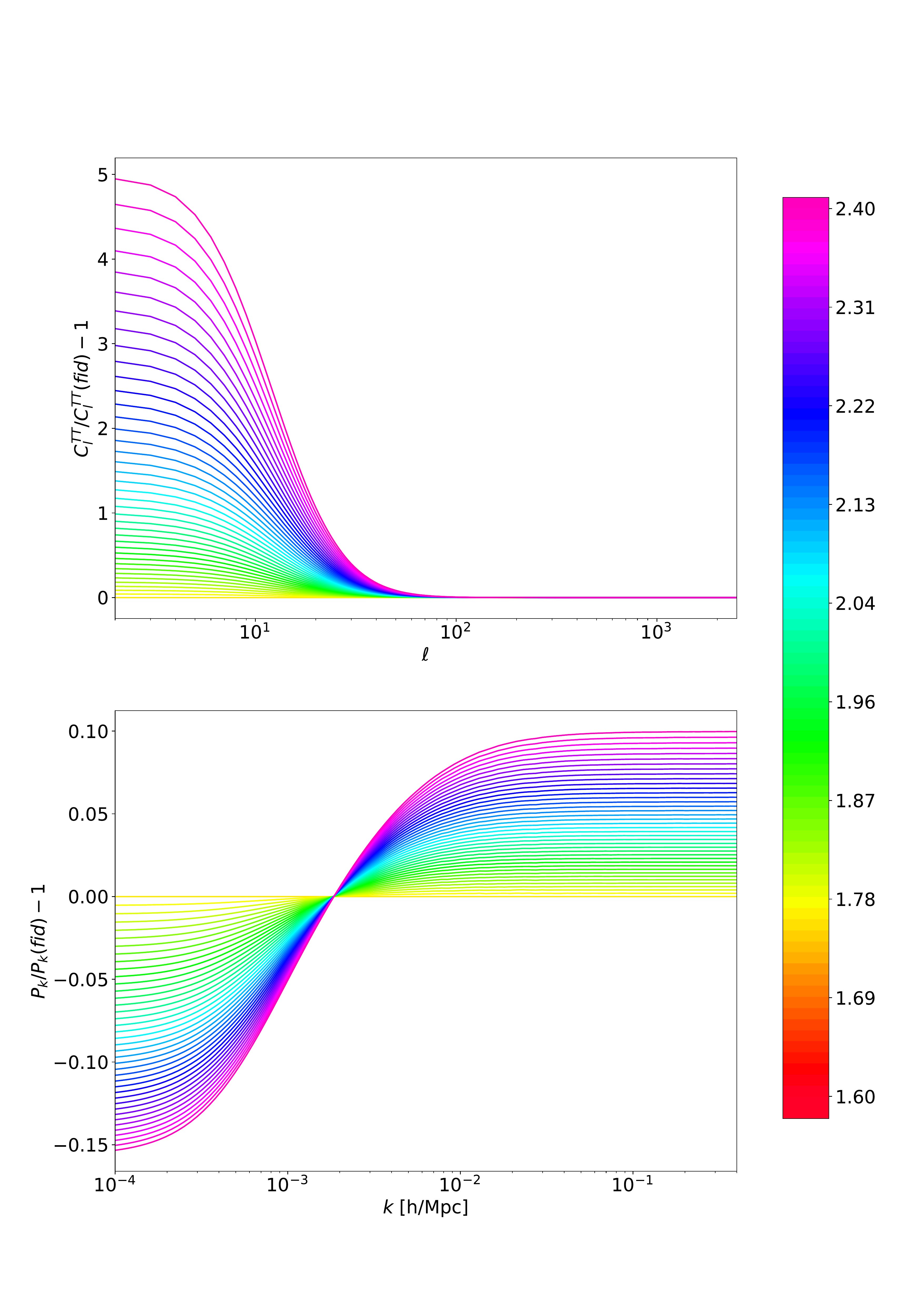}
   \caption{We show the relative deviation of the CMB temperature-temperature (top panel) and the matter (bottom panel) power spectra w.r.t.~a fiducial model for different values of $\alpha_{\rm B}\left(a=1\right)$ (the corresponding color for each value is shown on the color bar). We chose to pick models with both $\alpha_{\rm B}<2$ and $\alpha_{\rm B}>2$, to see if some kind of discontinuity could be detected. We notice that the spectra seems continuous and smooth at this point.}
  \label{fig:aB2}
\end{figure}

\noindent {\it Discontinuity in the evolution equations}: For general $\alpha_{\rm B}$, one can straightforwardly derive the (coupled) evolution equations for $\delta\psi_M$ and $\Phi$. Using these and taking the limit as $\alpha_{\rm B} \to 2$, one recovers the constraint, Eq.~\eqref{eq:aB2constraint}, from the equation of motion for $\delta\psi_M$ and can then use this constraint to solve for $\delta\psi_M$, arriving at a single second order evolution equation for $\Phi$. This reads
\begin{align}
&\Phi^{\prime\prime} + \left(2 \mathcal{H} + \frac{2 a^2 V_{\psi_M}}{\psi_M^{\prime}}\right) \Phi^{\prime} + k^2 \Phi = 0,
 \label{eq:eomsPhippLim}
\end{align}
where we have again assumed that there is no cosmological running of the Planck mass and will also assume that the speed of gravitational waves is precisely the speed of light in what follows -- both assumptions are met for the models considered in this paper and violating them would complicate the expressions shown here, although not the qualitative conclusions of this appendix. Now suppose we instead first set $\alpha_{\rm B} = 2$ in the full quadratic action and then derive the residual evolution equation for the remaining degree of freedom. Again we recover the constraint, Eq.~\eqref{eq:aB2constraint}, this time from a Lagrange multiplier in the quadratic action. However, the evolution equation for $\Phi$ now instead reads 
\begin{align} \label{eq:eomsPhipp2}
\Phi^{\prime\prime} + \left(2 \mathcal{H} + \frac{2 a^2 V_{\psi_M}}{\psi_M^{\prime}}\right) \Phi^{\prime}
+ k^2 \Phi & \nonumber \\
-  \frac{\Phi}{2} \left(\left(6 + \hat{\alpha}_\textrm{K}\right) \mathcal{H}^2 + \frac{\varphi^{\prime}{}^2}{M_P^2}\right) 
 &= 0.
\end{align}
This is identical to Eq.~\eqref{eq:eomsPhippLim}, except for the addition of the last term. While the last term is suppressed with respect to the second last in the sub-horizon limit, this nevertheless again hints at evolutions crossing the $\alpha_{\rm B} = 2$ point being ill-defined, since there does not seem to be a uniquely defined evolution across this point. However, carrying out the analogous calculation in Newtonian gauge gives the same equations up to terms proportional to $\alpha_{\rm B}^\prime$, which also vanish in this limit.
This suggests that the above-mentioned discontinuity in the evolution equations might be a gauge artefact \cite{Ijjas:2017pei}.
\\

Summarising, both these issues are alarming and should be investigated in more detail. However, a definitive answer can be given only after further investigation, and this is beyond the scope of this paper.
Despite the above issues, it is important to notice -- as shown in Fig.~\ref{fig:aB2} -- that the CMB and matter spectra do not show any discontinuity when crossing $\alpha_{\rm B} = 2$. This shows that it is possible to solve this system in such a way that they do not show any observable discontinuity at $\alpha_{\rm B} = 2$. This corresponds to the second case considered above, Eq.~\eqref{eq:eomsPhipp2}. In addition, given that the majority of the models considered and consistent with current observational constraints never crosses $\alpha_{\rm B} = 2$, a hard bound at this point should not affect our results qualitatively. While these observations do not resolve the above issues as such, they are nevertheless encouraging and suggest that they may be resolved without invalidating other parts of the analysis. For this reason we here put these issues to one side and do evolve across $\alpha_{\rm B} = 2$ in this way.

\bibliography{priors_draft.bib}

\end{document}